\documentclass[preprint]{aastex631}

\usepackage{hyperref}
\usepackage{lipsum}
\usepackage{graphicx}
\usepackage{enumitem}
\usepackage{float}
\usepackage{listings}
\usepackage{tabularx}
\usepackage{booktabs}
\usepackage{verbatim}

\newcommand{\package}[1]{\texttt{#1}}

\newcommand{\update}[1]{\textcolor{black}{#1}}
\newcommand{\updatetwo}[1]{\textcolor{black}{#1}}
\newcommand{\updatethree}[1]{\textcolor{black}{#1}}

\newcommand{\degree}{$^{\circ}$}

\shorttitle{LSST Centaur Discoveries}
\shortauthors{Murtagh et al.}

\graphicspath{{./}{figures/method/}{figures/results/}}
\begin{document}

\defcitealias{gladman08}{G08}
\defcitealias{sarid19}{S19}

\title{Predictions of the LSST Solar System Yield: Discovery Rates and Characterizations of Centaurs}

\author[0000-0001-9505-1131]{Joseph Murtagh}
\affiliation{Astrophysics Research Centre, School of Mathematics and Physics, Queen's University Belfast, BT7 1NN, UK}
\correspondingauthor{Joseph Murtagh}
\email{jmurtagh05@qub.ac.uk}

\author[0000-0003-4365-1455]{Megan E. Schwamb}
\affiliation{Astrophysics Research Centre, School of Mathematics and Physics, Queen's University Belfast, BT7 1NN, UK}

\author[0000-0001-5930-2829]{Stephanie R. Merritt}
\affiliation{Astrophysics Research Centre, School of Mathematics and Physics, Queen's University Belfast, BT7 1NN, UK}

\author[0000-0003-0743-9422]{Pedro H. Bernardinelli}
\affiliation{DiRAC Institute, University of Washington, 3910 15th Ave NE, Seattle, WA, USA}
\affiliation{Department of Astronomy, University of Washington, 3910 15th Ave NE, Seattle, WA, USA}

\author[0009-0005-5452-0671]{Jacob A. Kurlander}
\affiliation{Department of Astronomy, University of Washington, 3910 15th Ave NE, Seattle, WA, USA}

\author[0000-0002-0672-5104]{Samuel Cornwall}
\affiliation{Department of Aerospace Engineering, University of Illinois at Urbana-Champaign, 104 S Wright St, Urbana, IL, 61801, USA}

\author[0000-0003-1996-9252]{Mario Juri\'{c}}
\affiliation{DiRAC Institute, University of Washington, 3910 15th Ave NE, Seattle, WA, USA}
\affiliation{Department of Astronomy, University of Washington, 3910 15th Ave NE, Seattle, WA, USA}

\author[0000-0002-8418-4809]{Grigori Fedorets}
\affiliation{Finnish Centre for Astronomy with ESO, University of Turku, FI-20014 Turku, Finland}
\affiliation{Department of Physics, University of Helsinki, PO Box 64, 00014, Helsinki, Finland}

\author[0000-0002-1139-4880]{Matthew J. Holman}
\affiliation{Center for Astrophysics $\mid$ Harvard \& Smithsonian, 60 Garden St., MS 51, Cambridge, MA 02138, USA}

\author[0000-0002-1398-6302]{Siegfried Eggl}
\affiliation{Department of Aerospace Engineering, University of Illinois at Urbana-Champaign, 104 S Wright St, Urbana, IL 61801, USA}
\affiliation{Department of Astronomy, University of Illinois at Urbana-Champaign, 104 S Wright St, Urbana, IL 61801, USA}

\author[0000-0002-4547-4301]{David Nesvorn\'{y}}
\affiliation{Department of Space Studies, Southwest Research Institute, 1301 Walnut St., Suite 400, Boulder, CO 80302, USA}

\author[0000-0001-8736-236X]{Kathryn Volk}
\affiliation{Planetary Science Institute, 1700 East Fort Lowell, Suite 106, Tucson, AZ 85719, USA}

\author[0000-0001-5916-0031]{R. Lynne Jones}
\affiliation{Rubin Observatory, 950 N. Cherry Ave., Tucson, AZ 85719, USA}
\affiliation{Aston Carter, Suite 150, 4321 Still Creek Dr., Burnaby, BC V5C6S, Canada}

\author[0000-0003-2874-6464]{Peter Yoachim}
\affiliation{DiRAC Institute, University of Washington, 3910 15th Ave NE, Seattle, WA, USA}
\affiliation{Department of Astronomy, University of Washington, 3910 15th Ave NE, Seattle, WA, USA}

\author[0000-0001-5820-3925]{Joachim Moeyens}
\affiliation{Asteroid Institute, a program of B612 Foundation, 20 Sunnyside Ave., STE F, Mill Valley, CA 94941, USA}
\affiliation{DiRAC Institute, University of Washington, 3910 15th Ave NE, Seattle, WA, USA}
\affiliation{Department of Astronomy, University of Washington, 3910 15th Ave NE, Seattle, WA, USA}

\author[0009-0009-2281-7031]{Jeremy Kubica}
\affiliation{McWilliams Center for Cosmology, Department of Physics, Carnegie Mellon University, Pittsburgh, PA 15213, USA}
\affiliation{LSST Interdisciplinary Network for Collaboration and Computing Frameworks, 933 N. Cherry Avenue, Tucson AZ 85721}

\author[0000-0001-6984-8411]{Drew Oldag}
\affiliation{DiRAC Institute, University of Washington, 3910 15th Ave NE, Seattle, WA, USA}
\affiliation{Department of Astronomy, University of Washington, 3910 15th Ave NE, Seattle, WA, USA}
\affiliation{LSST Interdisciplinary Network for Collaboration and Computing Frameworks, 933 N. Cherry Avenue, Tucson AZ 85721}

\author[0009-0003-3171-3118]{Maxine West}
\affiliation{DiRAC Institute, University of Washington, 3910 15th Ave NE, Seattle, WA, USA}
\affiliation{Department of Astronomy, University of Washington, 3910 15th Ave NE, Seattle, WA, USA}
\affiliation{LSST Interdisciplinary Network for Collaboration and Computing Frameworks, 933 N. Cherry Avenue, Tucson AZ 85721}

\author[0000-0001-7335-1715]{Colin Orion Chandler}
\affiliation{DiRAC Institute, University of Washington, 3910 15th Ave NE, Seattle, WA, USA}
\affiliation{Department of Astronomy, University of Washington, 3910 15th Ave NE, Seattle, WA, USA}
\affiliation{LSST Interdisciplinary Network for Collaboration and Computing Frameworks, 933 N. Cherry Avenue, Tucson AZ 85721}



\begin{abstract}
The Vera C. Rubin Observatory Legacy Survey of Space and Time (LSST) will start by the end of 2025 and operate for ten years, offering billions of observations of the southern night sky. One of its main science goals is to create an inventory of the Solar System, allowing for a more detailed understanding of small body populations including the Centaurs, which will benefit from the survey's high cadence and depth. In this paper, we establish the first discovery limits for Centaurs throughout the LSST's decade-long operation using the best available dynamical models. Using the survey simulator \texttt{Sorcha}, we predict a \update{$\sim$7-12 fold} increase in Centaurs in the Minor Planet Center (MPC) database, reaching \update{$\sim$1200-2000} (dependent on definition) by the end of the survey - about \update{50\%} of which are expected within the first \update{2} years. Approximately \update{30-50} Centaurs will be observed twice as frequently as they fall within one of the LSST's Deep Drilling Fields (DDF) for on average only up to two months. Outside of the DDFs, Centaurs will receive $\sim$200 observations across the \textit{ugrizy} filter range, facilitating searches for cometary-like activity through PSF extension analysis, as well as fitting light-curves and phase curves for color determination. Regardless of definition, over \update{200} Centaurs will achieve high-quality color measurements across at least three filters in the LSST's six filters. These observations will also provide over \update{300} well-defined phase curves in the \textit{griz} bands, improving absolute magnitude measurements to a precision of 0.2 mags.
\end{abstract}



\section{Introduction} \label{sec:intro}

The Centaurs are a class of small, icy bodies that orbit the Sun on giant-planet-crossing paths. They are a transient population, evolving inwards into the solar system from the trans-Neptunian population due to frequent gravitational perturbations with the giant planets, leading to orbits with dynamical timescales on the order of $\sim$1-10 Myr \citep{tiscareno03, disisto07, volk08, bailey09, disisto20}. Centaurs face a variety of fates from these interactions; some may have impacts with the giant planets, whilst some may be ejected out of the Solar System entirely \citep{dones15}. Others still may diffuse inwards from the trans-Neptunian region into the solar system to become short period comets, such as the Jupiter-family comets (JFCs) \citep{holman93, duncan97, levison97, duncan04, emelyanenko05, volk08, jewitt09, sarid19, guilbertlepoutre23}. Further evidence of this evolutionary continuum is seen in the color distribution of known Centaurs following closely with that of the smaller sized trans-Neptunian objects (TNOs) \citep{tegler03, peixinho03, tegler08, peixinho12, fraser12, tegler16, wong16, wong17}, whilst also having a size distribution more similar to that of the JFCs \citep{sheppard00, jedicke02, bauer13, fernandez13}. \update{Study of Centaur properties in ensemble is therefore a means of providing insight into the evolution of dynamically scattering TNOs into present day comets, as well as probing the evolution of how their surfaces are processed.}

\update{There is no unanimous definition of a Centaur in literature, with cuts being made in orbital space and/or in dynamical timescales \citep[e.g.][]{levison97, tiscareno03, disisto07, jewitt09, sarid19, peixinho20}, however the most common community consensus places their orbits between Jupiter and Neptune. It is due to this variety of definitional cuts, as well as their inherent dynamical instability that Centaurs display, that there is no consistent count of either the observed or intrinsic number of Centaurs.} Recently, \cite{volk24} applied the \cite{gladman08} definition to all known multi-opposition outer solar system objects within the Minor Planet Center\footnote{\href{https://minorplanetcenter.net/iau/lists/MPLists.html}{https://minorplanetcenter.net/iau/lists/MPLists.html}} resulting in a list of 168 known Centaurs at the time of writing. The \update{relatively small number of known Centaurs (compared to other TNO populations)} is owed partly due to \update{there having been a lack of} dedicated Centaur discovery surveys \update{(see \citealt{kurlander24} for a survey of the Pan-STARRS1 detection catalog)}, with the majority being serendipitously \update{discovered} in other TNO, Main Belt asteroid (MBA), or Near Earth Object (NEO) searches \citep[e.g.][]{petit08, trujillo08, sheppard11, rabinowitz12, adams14, weryk16, sheppard16, bannister18}. Lack of follow-up observations, or detection algorithms being designed for either slow or fast movers in these surveys means that the Centaurs, who by any definition cover a wide range of orbital space, are less preferentially observed. Their small inherent population size relative to the TNOs or asteroids adds to this issue of missing Centaur discoveries. A survey that probes this region of the outer solar system will thus be required to be designed to cover these gaps in observation space.

At the Vera C. Rubin Observatory in Chile, the Legacy Survey of Space and Time (LSST) is scheduled to begin survey operations by the end of 2025. With a 9.6 deg$^{2}$ field of view, one of the LSST's science goals will be cataloging the entire solar system \citep{ivezic19}. The LSST is set to revolutionize solar system study - with its cadence of 30s exposures covering 18,000 deg$^2$ every three nights across six broad-band \textit{ugrizy} filters \citep{lsst09, ivezic19, bianco22}, the LSST has been uniquely tuned to give the best compromise of observing strategy across all small body populations. Specific patches of the sky known as Deep Drilling Fields (DDFs) will be targeted with higher temporal cadences, allowing for observations at a stacked magnitude deeper than the main Wide-Fast-Deep (WFD) survey over 10 years \citep{bianco22}, whilst mini-surveys will be carried out in areas like the Northern Ecliptic Spur (NES, +10$^{\circ}$ ecliptic latitude), wherein additional \textit{griz} observations in the ecliptic plane will allow for enhanced detections of outer solar system objects with longer orbital periods \citep{schwamb23}. All of these combined mean that over its planned 10 year long observational baseline, it has been predicted that the LSST will discover roughly an order of magnitude more objects in each small body population residing in the solar system \citep{jones09, lsst09, solontoi10, shannon15, grav16, silsbee16, veres17, jones18, ivezic19, fedorets20, hoover22}. The LSST's deep, high cadence observations will provide particular opportunities for investigations into the key areas of Centaur characteristics, including probing potential cometary activity, ring systems, and developing quality phase curves, rotational light curves, and photometric colors \citep{jones09, lsst09, schwamb18, schwamb21, schwamb23}. 

\update{Despite its readily apparent strengths for Centaur discovery}, there have been no estimates \update{on the Centaur yield} within the LSST. The recent cadence optimization work by \cite{schwamb23} specifically excludes the Centaur population in their simulations, as does the \cite{lsst09} \update{analysis}. \update{Predictions for the Centaur discovery metrics, including the total number of observations available per object, when they are discovered, and how many will be discovered, are vitally important in understanding the potential for Centaur science within the LSST through light curve, phase curve, and surface color studies. Further, such predictions for Centaur observations are crucial for understanding the gaps in the LSST's observation cadence, which will enable the design of follow-up observational campaigns to supplement and bolster the LSST, such as for investigating Centaur activity.} In this work, we \update{address this gap in planning for Centaur science within the LSST by providing the very} first estimates of Centaur \update{discovery metrics} using the current best dynamical and physical models from literature. 

\update{In Section \ref{sec:2}, we describe our methods of modeling discovery, using the survey simulator \texttt{Sorcha}, a solar system survey simulator, simulated LSST observation cadences, and a model for the Centaurs. In Section \ref{sec:3} we outline the results of these simulations, including discovery rates for Centaurs within the first few years of the LSST, potentials for Centaur activity probing, and phase curve, light curve, and surface color metrics. Finally, Section \ref{sec:4} summarizes the major results of Centaur discovery and characterization within the LSST, and also highlights the potential limitations to this work.}

\section{Methods}\label{sec:2}

In this work we use the leading dynamical model from \cite{nesvorny19} of the Centaur population, calibrated Centaur detections in the Outer Solar System Origins Survey (OSSOS) \citep{bannister18}, in order to gain the first predictions for the number of Centaurs that will be discoverable within the LSST. From these predicted observations, we explore the analysis that is possible within both early and later years of operation of the LSST through the measurements of light curves, phase curves, color information, and any cometary-like activity. As the LSST is a well-characterized survey (i.e. it's depth per observation, pointing histories, and detection efficiencies will be measured; see \cite{lawler18b} for further discussion), we simulate what objects would be detected within the LSST, and forward bias them in order to estimate the number of LSST Centaur discoveries. 

\subsection{Simulating with Sorcha}\label{sec:2.1}
We simulate our Centaur discoveries within the LSST using \texttt{Sorcha} \citep{merritt25, holman25}, an open-source, modular, \texttt{Python} survey simulator designed with surveys such as the LSST in mind. We refer the reader to \cite{merritt25} for a full discussion on how \texttt{Sorcha} has been designed, but we highlight the basic functionality here for clarity. \texttt{Sorcha} takes an input model of the Centaurs, described by object orbital elements and physical parameters including absolute magnitude in $r$ band, photometric colors with respect to $r$, and the same phase curve parameters in all bands (see Section \ref{sec:2.3}). Ephemerides are generated for objects using the in-built N-body integrator ASSIST \citep{holman23} - itself an extension of the REBOUND package \citep{rein12, rein15}, using its IAS15 integrator (Gauss-Radau integrator with Adaptive Step-size control, 15th order \citealt{rein15}) with the Sun, Moon, planets, and 16 massive asteroids and dwarf planets (see Table 5 in \cite{merritt25} for list) as perturbers. Given a database of the LSST's pointings (see Section \ref{sec:2.2}), \texttt{Sorcha} then computes if the object is located within a 2.26\degree radius of the pointing center, and then applies the LSSTCam footprint \citep{lsst09, ivezic19}. For those objects that are not located within a gap in the CCD chips, a magnitude is calculated using the absolute magnitude (see Section \ref{sec:2.3.2}) and the phase angle of the observation (see Section \ref{sec:2.3.4}). \updatethree{Any sources brighter than $m_r = 16$  are removed as they exceed the estimated saturation threshold of the LSST \citep{ivezic19}}. The source detection efficiency is modeled for all observations by a modified sigmoid function from \cite{chelsey17} as follows:

\begin{equation} \label{eq:fade}
    \epsilon(m_{PSF}) = \frac{F}{1 + e^\frac{m_{PSF}-m_{5\sigma}}{w}}
\end{equation}
where $\epsilon(m_{PSF})$ represents the probability of detection, $F$ is the survey's peak detection efficiency, $m_{PSF}$ and $m_{5\sigma}$ are (respectively) the object's point spread function (PSF) magnitude \citep[the source magnitude measured by the Rubin source detection algorithm;][]{merritt25} and $5\sigma$ limiting magnitude of the observation at the source's location on the camera focal plane, and $w$ is the width of the function. In this work we select $w$ = 0.1 following the Sloan Digital Sky Survey (SDSS) \citep{annis14}, and $F$ = 1 \citep[source detection for the LSST has already been well-tuned for bright sources, and the true value is predicted to be close to 1;][]{juric21}. Objects which have an apparent magnitude in the \textit{r} band $m_r < 16$ are removed as they are below the estimated saturation limit for the LSST \citep{ivezic19}. Finally, the LSST will also be the first survey at this scale with a dedicated moving object detection pipeline designed to search inward of Venus to beyond Neptune \citep{ivezic19, juric21}. \texttt{Sorcha} models the Rubin Solar System Processing (SSP) object linking algorithm, which will associate moving sources through the linking of 3 pairs of nightly observations (or, tracklets) which are separated spatially $> 5"$ and temporally $< 90$ mins, all within a 15 day window, with a design specification of 95\% of discovery opportunities that meet these criteria being successfully linked \citep{lsst09, ivezic13, myers13, ivezic19, juric20}. The resulting output from \texttt{Sorcha} contains predicted observations of all input objects that would be detectable in the LSST. 

\updatethree{Simple magnitude and visibility cuts can approximate the number of detections per object but fail to capture the full complexity of a real survey. These cuts assume uniform observation conditions, which do not apply to the LSST's observing cadence. Using a survey simulator with ephemeris generation accounts for object motion and the ability to link detections across multiple visits. Additionally, the simulator can incorporate detection efficiency, allowing for the possibility of detecting objects fainter than the nominal survey magnitude limit through repeated observations. These effects are difficult or impossible to model accurately with simple cuts and require a survey simulator for realistic results.}


\subsection{The LSST Cadence Simulation}\label{sec:2.2}
\begin{figure*}[!h]
    \centering
    \includegraphics[width=\textwidth]{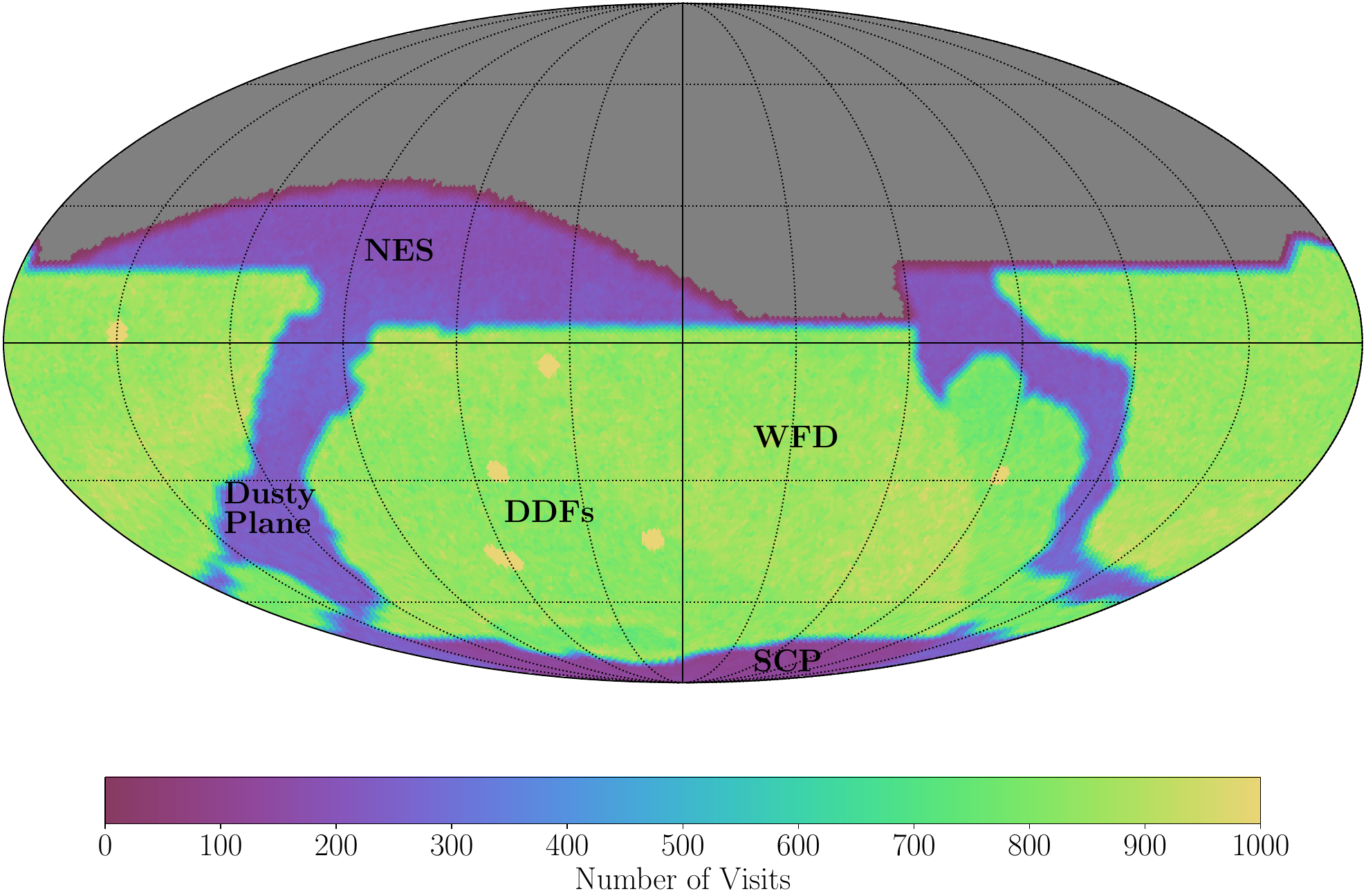}
    \caption{Skymap of the number of visits across all filters of the 10 year LSST survey cadence based on the one snap v4.0 simulation \citep{scocv3}. \update{The main Wide-Fast-Deep (WFD) survey making up $\sim$80\% of the total survey time will receive $\sim$800 visits per pointing. There are additional "Mini" surveys highlighted that make up $\sim$3-10\% of the survey time. These cover differing areas of the sky, including the Northern Ecliptic Spur (NES), the South Celestial Pole (SCP), and the Dusty Plane. Additionally highlighted are Deep Drilling Fields (DDFs); regions within the WFD that will receive deeper coverage and more frequent temporal sampling and use a total of $\sim$6.5\% of the survey time. For complete detail of the survey strategy, see \cite{scocv3}}}
    \label{fig:nvisits}
\end{figure*}

We use the output baseline cadence from the v4.0 of the observing strategy \citep{scocv3} throughout all of our runs, with the number of visits across all filters shown in skymap form in Figure \ref{fig:nvisits}. \updatetwo{This simulation is generated by the \package{rubin\_sim} package \citep{bianco22, yoachim23} and the Rubin Observatory scheduler, \package{rubin\_scheduler} \citep{naghib19, yoachim24b}\footnote{For the most up-to-date cadence simulations, see \hyperlink{https://s3df.slac.stanford.edu/data/rubin/sim-data/}{https://s3df.slac.stanford.edu/data/rubin/sim-data/}}. It is based on a model observatory of the LSST that calculates on-sky limiting magnitudes and seeing conditions at each pointing based on assumed realistic weather conditions, telescope and camera performance, and individual filter responses with up-to-date mirror coating specifications \citep{connolly14, delgado14, delgago16, yoachim16, lsst17, jones18, jones20, naghib19, bianco22}.} The most up-to-date observing cadence as determined by the Rubin Survey Cadence Optimization Committee \citep[SCOC;][]{scocv1, scocv2, scocv3}, including exposure time per filter and on-sky pointing position, is then applied, allowing simulated observations to be forward biased to what the LSST will `see'. \updatetwo{Compared to prior cadence simulations, more time is spent on engineering downtime in year 1. This is, however, largely gained back in subsequent years and so only affects year 1 metrics.} Whilst this simulation starts in May 2025, Rubin Operations have assumed a need for at minimum 6 weeks worth of contingency and construction delay \citep{esp}\footnote{See \href{https://rubinobservatory.org/about/construction}{https://rubinobservatory.org/about/construction} for the most recent timeline updates.}, so true start dates will differ. 

We note two different versions of this baseline - \package{one\_snap\_v4.0} wherein the \textit{grizy} observations are obtained in a single 29.2s exposure (referred to as one snap), and \package{baseline\_v4.0} with 2x15s exposures (referred to as two snap). In either case, the \textit{u} exposure is set at a single 38s exposure \citep{scocv3}. We focus in the main body of this work on the one snap simulation, as the cadence recommendations are to move operations towards this as the feasibility of cosmic ray rejection is tested during commissioning. 

\subsection{Centaur Model}\label{sec:2.3}
\update{Initially we construct a Centaur} model based on the dynamically-driven definition from \cite{gladman08} (hereafter referred to as \citetalias{gladman08}), which makes cuts in perihelion distance \textit{q} $>$ 7.35 au, Tisserand parameter with respect to Jupiter $T_J >$ 3.05, and semimajor axis \textit{a} $<$ 30.1 au \updatetwo{(as well as implicitly an eccentricity $e <$ 0.756)}. \update{Due to the lack of community consensus on Centaur orbital definition,} we also explore two alternative definitions for Centaurs to investigate the effects \update{that differing population coverage in orbital space} has on \update{overall} detections. We look at the definition set out in \cite{sarid19} (hereafter referred to as the \citetalias{sarid19} sample), namely \textit{q} $>$ 5.2 au and aphelion distance \textit{Q} $<$ 30.1 au \updatetwo{(and $e <$ 0.71)} - a definition that places objects entirely within the giant planet region, giving insight into the transition objects not accounted for within the \citetalias{gladman08} sample, but excluding eccentric objects with aphelia more distant than Neptune's orbit. Finally, we also employ a hybrid definition (hereafter referred to as the Hybrid sample), with \textit{q} $>$ 5.2 au and \textit{a} $<$ 30.1 au \updatetwo{(and $e < 0.83$)}, and no aphelia constraints - this has the benefit of including the base giant-planet crossing orbits that make up the bulk of the known Centaurs, whilst also including those on more eccentric, and so distant \update{perihelia} orbits, as well as those near transition objects included in the \citetalias{sarid19} sample.

\update{The resulting orbital distributions for all three models is shown in Figure \ref{fig:orbitalspace}. The \citetalias{gladman08} and Hybrid models appear the most similar due to the Hybrid otherwise containing only $\sim$150k low $q$ objects that do not appear in the \citetalias{gladman08} model. Conversely, the \citetalias{sarid19} model does not contain as many high semimajor axis objects, with a median value of $\sim$21 au.  All three broadly share the same eccentricity and inclination distributions, with the majority being concentrated in a central $e = 0.1-0.5$/$i = $ 0-50\degree range - approximately 1\% of each model is however also comprised of retrograde ($i >$ 90\degree) orbits.} All three definitions of these models are summarized in Table \ref{tab:orbitalspace}. The following sections detail how we \update{developed} the \citetalias{gladman08} model for the Centaurs, with Section \ref{sec:2.4} looking at the differences in this process needed to construct the alternative two models.

\subsubsection{Orbital Distribution}
\begin{figure*}
    \centering
    \includegraphics[width=\textwidth]{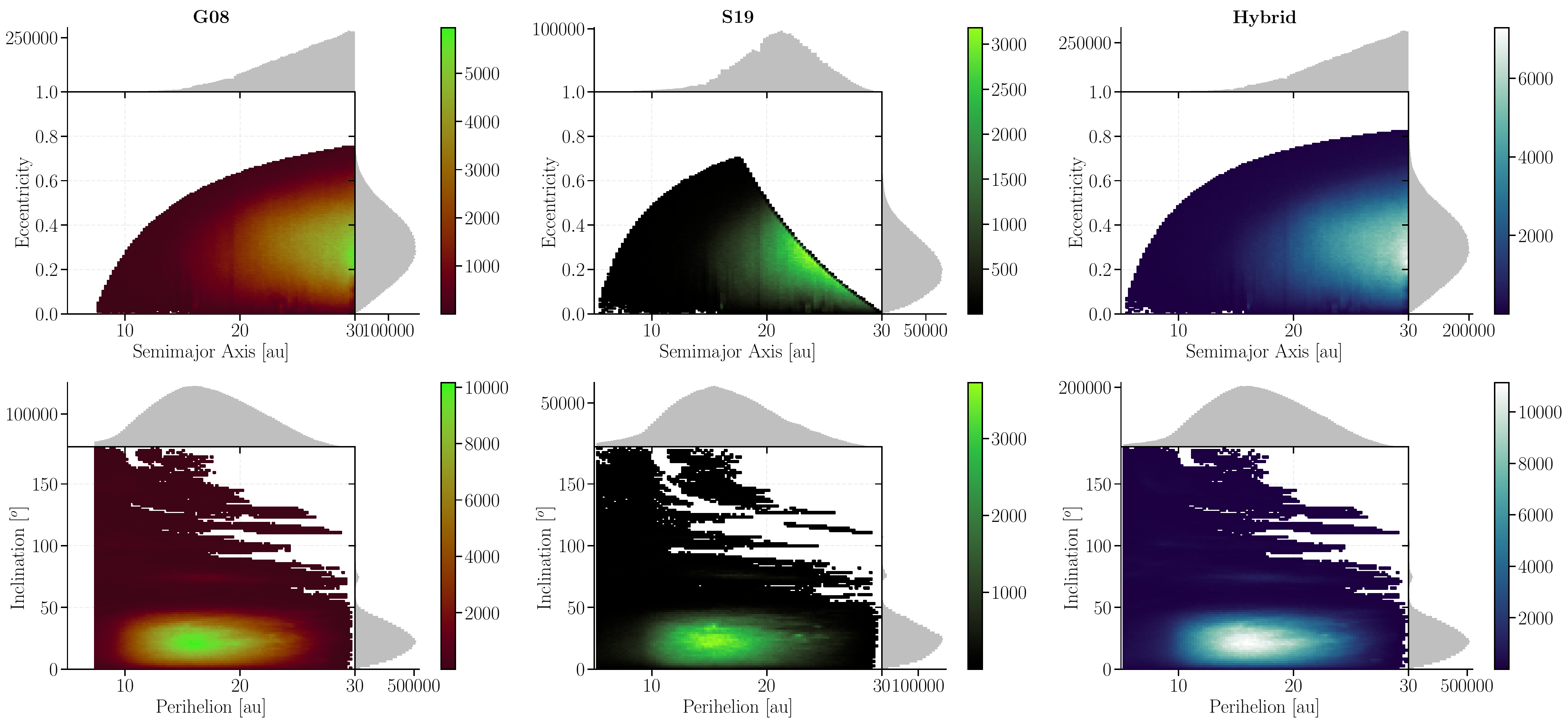}
    \caption{Orbital space 2D histograms for the three dynamical classifications of the input Centaur models used in this work; namely the \citetalias{gladman08} sample on the leftmost column, the \citetalias{sarid19} sample on the middle column, and the Hybrid sample in the rightmost column. The top row shows histograms of Centaur eccentricity as a function of semimajor axis, whereas the bottom row shows histograms of Centaur inclination as a function of perihelion distance.}
    \label{fig:orbitalspace}
\end{figure*}

We model \update{the} underlying Centaur population based on the end results of the N-body integration from \cite{nesvorny19}. \update{Here, a planetesimal disk of 10$^6$ particles (including 4000 Pluto sized objects below 30 au) was integrated with a grainy Neptune migration - here Neptune starts at a semimajor axis of 24 au and is simulated to present day with exponential e-folding timescales of $\tau_1$ = 30 Myrs and $\tau_2$ = 100 Myrs representing its migration at two  respectively. During the final Gyr of simulation, if an individual particle orbit reached a semimajor axis $a <$ 30 au, it was cloned 100 times via random small changes to their velocity vectors ($\delta V/V \sim 10^{-5}$), and saved with a $10^4$ yr cadence. After removing cometary and Trojan orbits and applying the \citetalias{gladman08} dynamical definition, this steady-state Centaur population of $\sim$2.6$\times10^7$ objects was then biased using the OSSOS survey simulator \citep{lawler18b}, and found to be consistent with actual OSSOS Centaur discoveries at the 1$\sigma$ level.} Orbital distributions are randomly drawn from this output in \textit{a, e,} and \textit{i} together, allowing for more than one object to have the same $a/e/i$ combination. \update{As the output angular elements are uncorrelated with each other, and in order to avoid clustering of the orbits,} the angular elements $\omega, \Omega,$ and \textit{M} are randomized U$\in$[0\degree, 360\degree) to account for the effects of orbital precession due to interactions with the giant planets from the original \cite{nesvorny19} simulation orbits. The resulting input orbital \update{distribution of the \citetalias{gladman08} model is} highlighted in Figure \ref{fig:orbitalspace}.


\begin{deluxetable}{cccc >{\color{black}}c}
\tablecaption{Dynamical definitions for each orbital distribution. \label{tab:orbitalspace}}
\tablewidth{1.0\columnwidth}
\tablehead{
\colhead{} & \colhead{\textit{q}} & \colhead{\textit{a} } & \colhead{\textit{Q}} & \colhead{\updatetwo{\textit{e}}} \\
\colhead{} & \colhead{[au]} & \colhead{[au]} & \colhead{[au]} & \colhead{} 
}
\startdata
\citetalias{gladman08} Sample & $>$ 7.35      & $<$ 30.1     &     ---     &    $<$ 0.756         \\
\citetalias{sarid19} Sample   & $>$ 5.2       &    ---      &   $<$ 30.1   &    $<$ 0.71         \\
Hybrid Sample                 & $>$ 5.2       & $<$ 30.1     &    ---      &    $<$ 0.83         \\    
\enddata
\end{deluxetable}

\subsubsection{Absolute Magnitude Distribution}\label{sec:2.3.2}
The physical size and albedo of an object will affect its brightness, and so how observable it will be. Including a physical size distribution in our model requires assumptions of albedo and color distributions which remain correlated \citep{alvarezcandal16, ayalaloera18, alvarezcandal19}, and are poorly constrained \update{for the Centaur population} with current samples. Instead, we opt to use a more easily modeled \update{$H_r$} distribution as it is related directly to the measured apparent magnitude. The apparent magnitude \textit{m} (ignoring rotational effects or activity - we assume our population is entirely inactive, see Section \ref{sec:3.3} for further discussion on the effects of activity) of an object at a given heliocentric distance \textit{r}, geocentric distance $\Delta$, and phase angle between Sun-Object-Observer $\alpha$ (whose effect on the scattering geometry is modeled by the phase function $\Phi$) is defined as:

\begin{equation} \label{eq:mag}
    m(\alpha, r, \Delta) = H + 5 \log_{10}(r \Delta) + \Phi(H, \alpha)    
\end{equation}

The size distribution of TNOs displays a `break' or a `knee' at diameters D$\sim$100 km, or an absolute magnitude $H_r$ = 7.7 \citep{shankman13, fraser14, lawler18}. As such they are \update{modeled} with a two-part power-law distribution, with different slope values being observed on either side of the break. \updatethree{As with \cite{nesvorny19} however, we opt to model the absolute magnitude distribution as a single power-law as defined by the following cumulative distribution:}

\begin{equation} \label{eq:N}
    N({\leq}H_r) = N_0 10^{\alpha_{0} (H_r - H_{0})}
\end{equation}
where the slope parameter $\alpha_{0}$ = 0.42, as per OSSOS observations \citep{shankman16, lawler18}. The value $N_0$ is a scaling factor representing the cumulative number of objects with an absolute magnitude $H_r \leq H_0$. \updatethree{This yields $\sim$33 objects with $H<7$, compared to roughly 20 objects from the knee power law in \cite{lawler18}. The difference of around 13 objects is less than 1\% of our final discovery results in Section \ref{sec:3.1}, and smaller than the random variation between runs.}

\begin{figure} 
    \centering
    \includegraphics[width=\columnwidth]{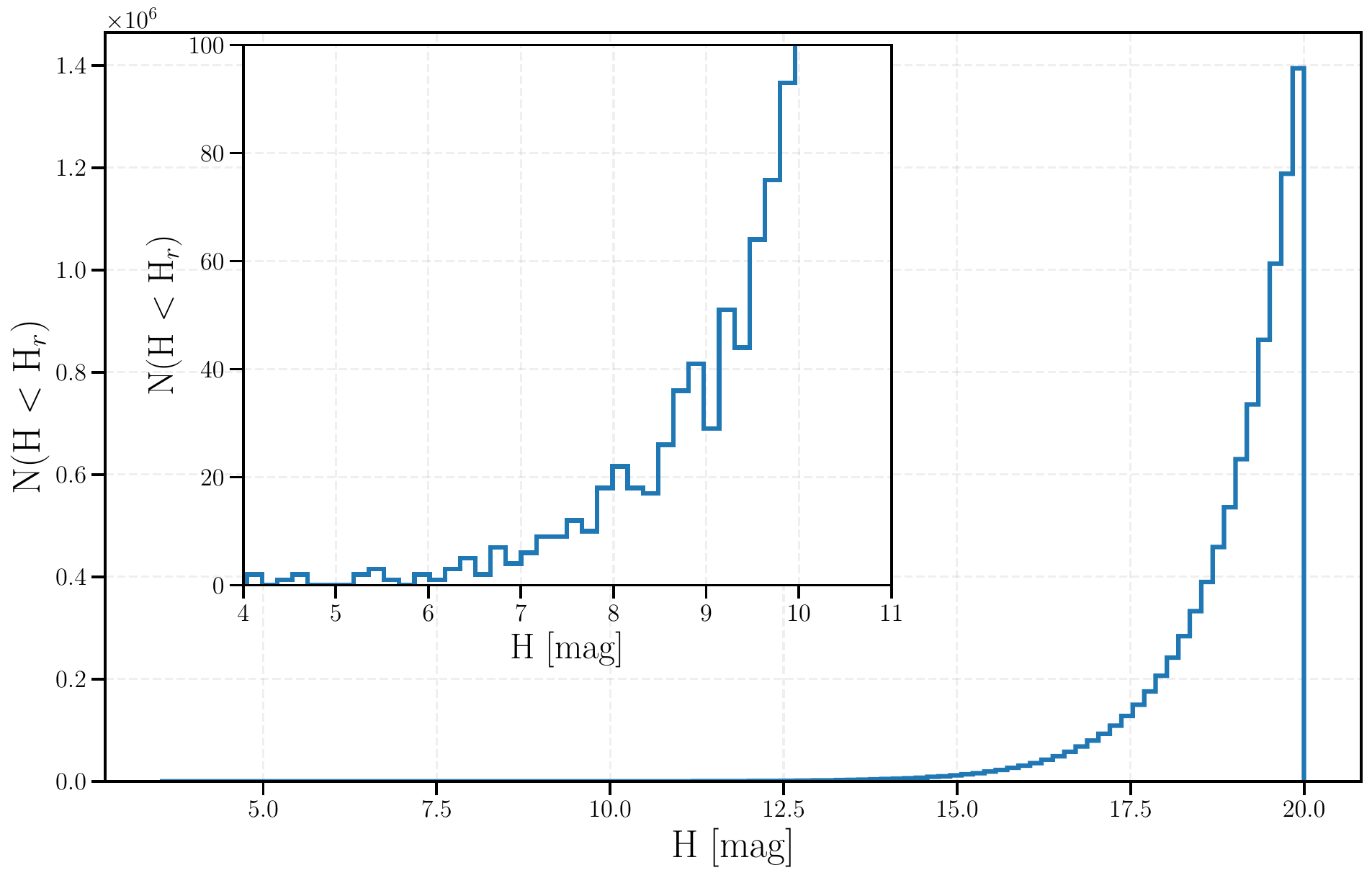}
    \caption{Cumulative histogram of the model \update{absolute magnitude \textit{H}} distribution of objects with \textit{H} less than a given absolute magnitude $H_r$ for the \citetalias{gladman08} sample. Inset represents a zoom-in of the bright end of the same distribution where $H <$ 11.}
    \label{fig:newhspace}
\end{figure}

\update{We adopt scaling constants} of \update{$N_0$ = 21,400} and $H_0$ = 13.7 for our \citetalias{gladman08} sample of Centaurs\update{, in line with recent estimates for the intrinsic \citetalias{gladman08} definition Centaurs from the debiased Pan-STARRS search of \cite{kurlander24}. We note that, within their stated population uncertainty of $^{+3400}_{-2800}$, this result is also consistent with the original debiased OSSOS population estimate of 21,000$\pm$8000 Centaurs with $H_r <$ 13.7 from \cite{nesvorny19}. We opt for the \cite{kurlander24} value over the \cite{nesvorny19} estimate however, due to it being a more well constrained estimate.} Each object to be simulated must be assigned an absolute magnitude from the distribution in Equation \ref{eq:N}. In order to find the maximum $H_r$ value that we need to simulate in our synthetic population, we first take the faintest 5$\sigma$ depth across all six \textit{ugrizy} bands recorded in the baseline cadence simulation ($\sim$26.2). Assuming an object is on a circular orbit with a perihelion distance of $\sim$5.2 au (with no phase effects), we then calculate an absolute magnitude using Equation \ref{eq:mag}. \updatetwo{From this, the faintest detectable object would have a $H_r \sim$ 19 - however the LSST limiting magnitude represents a detection efficiency of 50\% \update{(see Section \ref{sec:2.1}, Equation \ref{eq:fade})}. Objects which are slightly fainter will therefore still have a chance of being detected.} As such, assuming a detection efficiency function as detailed in \cite{merritt25}, we increase this value by a magnitude to $H_r$ = 20 in order to account for this effect. Combined with the previously discussed scaling factors of $\alpha$ and $N_0$, we find we need to simulate \textit{N}($H_r <$20) \update{$\sim$ 9.47$\times$10$^6$} Centaurs. We then uniformly sample \textit{N}($H_r <$20) number of orbits from our orbital model and assign them \textit{H} values via an inverse transform of Equation \ref{eq:N}, given by Equation \ref{eq:H} \updatetwo{and highlighted in Figure \ref{fig:newhspace}}. 

\begin{equation} \label{eq:H}
    H_r = \frac{1}{\alpha} \log_{10}\left(\frac{N(<H_r)}{N_0}\right) + H_{0}
\end{equation}
\updatetwo{(2060) Chiron is the largest Centaur ($H_V = 5.59$) reported in the MPC to date. Our $H$ distribution produces 8 objects larger than (2060) Chiron, of which 3 are brighter than apparent magnitudes of $m_r = 21.5$, and 1 lying within $\pm$10\degree of the ecliptic upon applying our orbital model. This is consistent with the observational constraints of Centaur discoveries from the Pan-STARRS Centaur survey in \cite{kurlander24}, who did not discover any new bright objects. It is also roughly consistent with previous shallow surveys that have contributed to the bulk of the MPC Centaury discovery catalog.}


\subsubsection{Colors}\label{sec:2.2.3}
\begin{figure*}
    \centering
    \includegraphics[width=\textwidth]{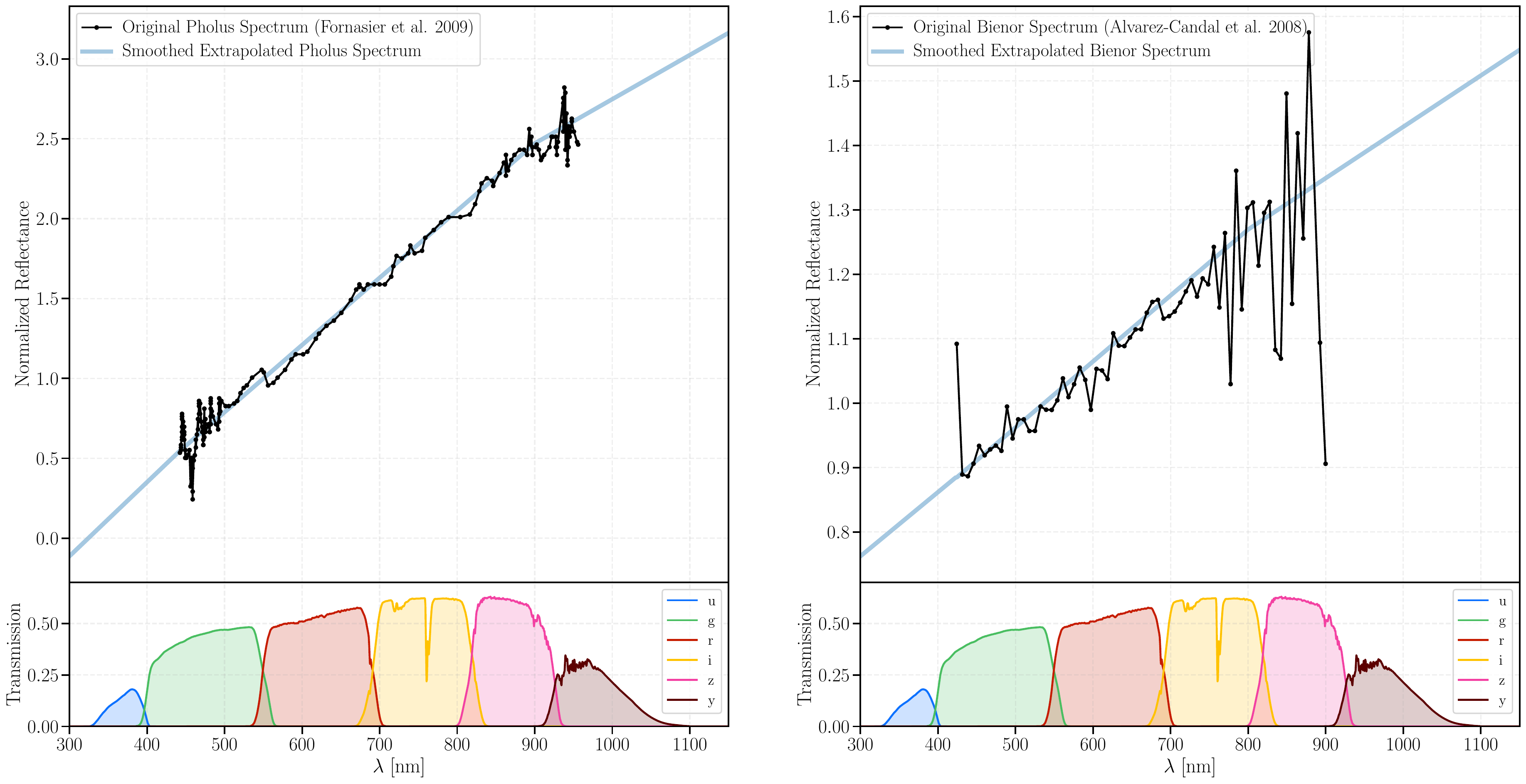}
    \caption{Original reflectance spectra of Pholus \citep[left,]
    []{fornasier09} and Bienor \citep[right,][]{alvarezcandal08}, normalized at 550nm. Overlaid are the new extrapolated, smoothed `spectra' that are used to convert to SED's and obtain color estimates. Below are the LSST filter throughputs for wavelength range reference. Note that Bienor's spectrum has been smoothed starting at $\sim$750nm in order to account for the increased noise in the proceeding range.}
    \label{fig:spectra}
\end{figure*}

The LSST will observe across 6 different filters, therefore we must account for the surface colors of Centaurs within our model population. Centaurs have been shown to display a bimodality in their color distribution, similar to that of the larger TNO population \citep{tegler98, tegler00, tegler03, peixinho03, barucci05, perna10, wong17}, owing in part to their shared dynamical history to the small hot TNO populations \citep{duncan97, levison97, volk08, fraser12, peixinho12}.  We use optical spectra of two canonical Centaurs as model objects; the `red' Centaur (5145) Pholus from \cite{fornasier09, perna10, barucci11} and `blue' Centaur (54598) Bienor from \cite{alvarezcandal08, guilbert09, perna10}. As Centaur and (small-sized) TNO spectra remain mostly featureless in the optical spectrum \citep{alvarezcandal08, barkume08, barucci08, fornasier09, barucci11, brown12b, barucci20}, we use the \package{curve\_fit} function from \package{scipy} \citep{virtanen20} to determine slopes at both red and blue ends of the existing spectra, and extending and smoothing over the noise at both ends to LSST filter wavelength ranges. In order to account for noise in the original spectra, we also smooth over the original data by applying a linear fit. The resulting three-component spectra are shown in Figure \ref{fig:spectra}. We multiply by the Solar spectrum \citep{kurucz05} again to convert these spectra to spectral energy distributions (SEDs), and then integrate the resulting flux under the LSST filter bandpasses via \package{phot\_utils} functions within \package{rubin\_sim}. The resulting colors are given in Table \ref{tab:cols}. We then use the blue:red color fraction of 3:1 as determined by \cite{wong17}, giving 75\% of our model population a Bienor-like color, and 25\% a Pholus-like color. This is however a first approximation for the population - in Section \ref{sec:3.1} we discuss the impact that our choice of color distribution has on discoverability.

\begin{deluxetable}{ccc}
\tablecaption{Estimated LSST Colors of Bienor and Pholus. \label{tab:cols}}
\tablewidth{\columnwidth}
\tablehead{
\colhead{LSST Color} & \colhead{Pholus (Red) Colors} & \colhead{Bienor (Blue) Colors} \\
\colhead{} & \colhead{[mags]} & \colhead{[mags]} 
}
\startdata
$u$-$r$ & 3.45                 & 1.86                  \\
$g$-$r$ & 1.00                 & 0.56                  \\
$i$-$r$ & -0.50                & -0.23                 \\
$z$-$r$ & -0.73                & -0.33                 \\
$y$-$r$ & -0.89                & -0.41                
\enddata
\end{deluxetable}

\subsubsection{Phase Curves}\label{sec:2.3.4}
In order to accurately calculate the apparent magnitude of each Centaur, phase effects must be accounted for. As a Centaur moves very little in its orbit during the LSST survey, the change in viewing geometry illuminating different surface fractions will not contribute considerably to their brightness changes - instead Centaur brightness variations are driven by the backscattering of light from surface particles \citep{rabinowitz07}. For Centaurs, this change in brightness tends to be linear \citep{buie92, rousselot05, bagnulo06, rabinowitz07, belskaya08, schaefer09, verbiscer13, fornasier14, ayalaloera18, dobson23}, with a slope measurable as $\beta$ using the following equation for the reduced magnitude (the apparent magnitude scaled to a geocentric and heliocentric distance of 1 au):

\begin{equation} \label{eq:linearmodel}
    M(\alpha) = H + \alpha \beta
\end{equation}
where \textit{H} is the absolute magnitude and $\alpha$ is the phase angle of the Sun-object-observer.

We apply a phase function to each Centaur by assigning each synthetic body a linear phase coefficient of $\beta$ = 0.071 mag deg$^{-1}$ - this is derived from the mean of the ensemble sample of 23 Centaurs and their measured linear phase coefficient from \cite{ayalaloera18}, \cite{alvarezcandal16}, and \cite{rabinowitz07}. The value of $\beta$ is itself a function of the photometric filter being observed in, however due to the small sample size of measured Centaur phase coefficients we uniformly assign this across all filters.  

\subsection{Alternative Centaur Models}\label{sec:2.4}
\begin{figure}
    \centering
    \includegraphics[width=\columnwidth]{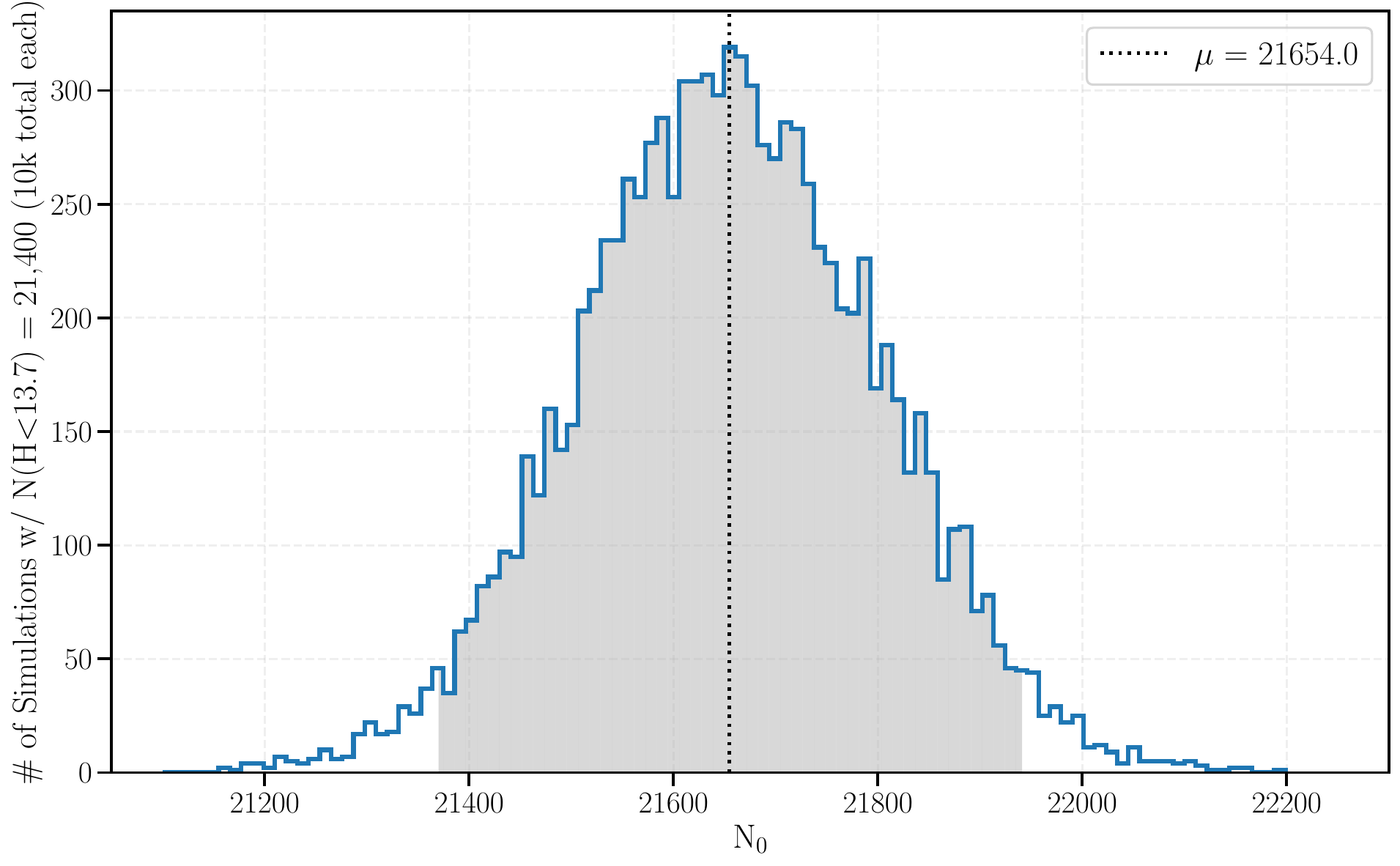}
    \caption{Histogram showing the number of 10,000 simulations of each trial $N_0$ value for the Hybrid definition that match the \citetalias{gladman08} sample scaling of \update{\textit{N}(\textit{H}$<$13.7) = 21,400. The dashed line represents the mean value of the distribution, and the shaded region is the 95\% confidence interval.}}
    \label{fig:N0grid}
\end{figure}

\update{With the \citetalias{gladman08} model now fully defined in terms of orbital distribution, absolute magnitude distribution, surface colors, and phase behavior, we next explore constructing \citetalias{sarid19} and Hybrid definition Centaur models to investigate how different Centaur orbital space classifications affect discovery predictions. The scaling parameter $N_0$ for the absolute magnitude distribution, which effectively controls the number of objects to be simulated, was calibrated to Pan-STARRS (in the case of \cite{kurlander24}) or OSSOS (in the case of \cite{nesvorny19}) detections of Centaurs that matched the \citetalias{gladman08} definition, and as such can not be applied to the \citetalias{sarid19} or Hybrid models. Definitionally, however, the \citetalias{gladman08} orbital space overlaps entirely with that of the Hybrid model. \updatetwo{The Hybrid sample may also be calibrated to OSSOS observations therefore by selecting the subset of orbits within the Hybrid model that match the \citetalias{gladman08} definition. The scaling parameter $N_0$ can then be tuned for these selected orbits until the same value of $N(<13.6)$ = 21,400 is obtained.} We trial a grid of $N_0$ values around the initial value of $N_0$ = 21,400, from 21,100 to 22,200 - for each value of $N_0$, Hybrid orbits are sampled from the original \cite{nesvorny19} model (as described in Section \ref{sec:2.3.2}) and the \citetalias{gladman08} orbits within selected. The cumulative count $N(<13.7)$ is then checked for this selected distribution against the \citetalias{gladman08} value of 21,400 - this is repeated $10^5$ times for each $N_0$, with a tally kept for each time the $N(<13.7)$ values match.} The resulting distribution of matched $N(<13.7)$ values is shown in Figure \ref{fig:N0grid} - from this the median value of \update{$N_0$ = 21,654 is selected and used} in Equation \ref{eq:N} for \update{the} Hybrid sample, resulting in \textit{N}($H_r <$20) \update{$\sim$ 9.58$\times$10$^6$} Centaurs to be simulated. As the Hybrid sample is now absolutely calibrated to OSSOS\update{/Pan-STARRS} detected Centaurs, \update{similarly the \citetalias{sarid19} orbits that overlap within the Hybrid orbits are selected} in order to obtain \update{the} calibrated \citetalias{sarid19} sample of \textit{N}($H_r <$20) \update{$\sim$ 3.13$\times$10$^6$} Centaurs. The process for obtaining colors and phase curve slopes for \update{both new models} are the same as outlined for the \citetalias{gladman08} sample.

\section{Results} \label{sec:3}

Throughout the following section, we present the results for the simulated \citetalias{gladman08}, \citetalias{sarid19}, and Hybrid Centaurs through \texttt{Sorcha} that pass the linking criteria outlined in Section \ref{sec:2.1}. All results utilize the one snap version of the LSST cadence simulation - the overall on-sky time is comparable between one snap and two snap simulations, only the distribution of when visits occur. As such, the following results remain consistent for both versions of the simulations.

\subsection{Discovery Yield}\label{sec:3.1}
\begin{figure*}
    \centering
    \includegraphics[scale=0.5]{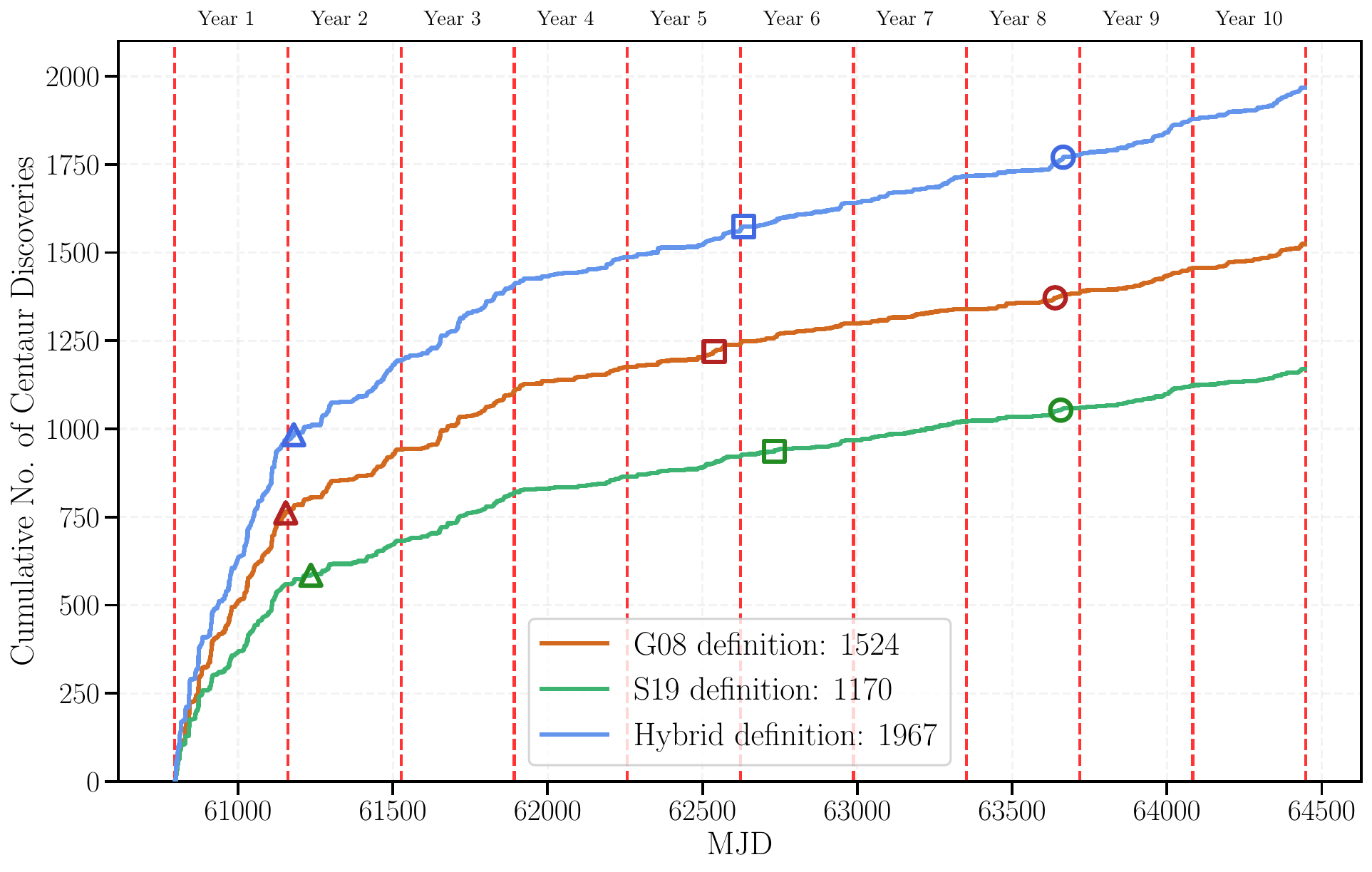}
    \caption{Cumulative histogram (with a bin size of one day) of the discovery rates of the different Centaur definitions over the 10 year LSST lifetime. Red dashed lines demark yearly boundaries for the survey, with labels above each block to highlight the year of operation. We highlight the points of 50\% (open triangle) 80\% (open square), and 90\% (open circle) completion for each population.}
    \label{fig:10yr}
\end{figure*}

\begin{figure*}
    \centering
    \includegraphics[width=\textwidth]{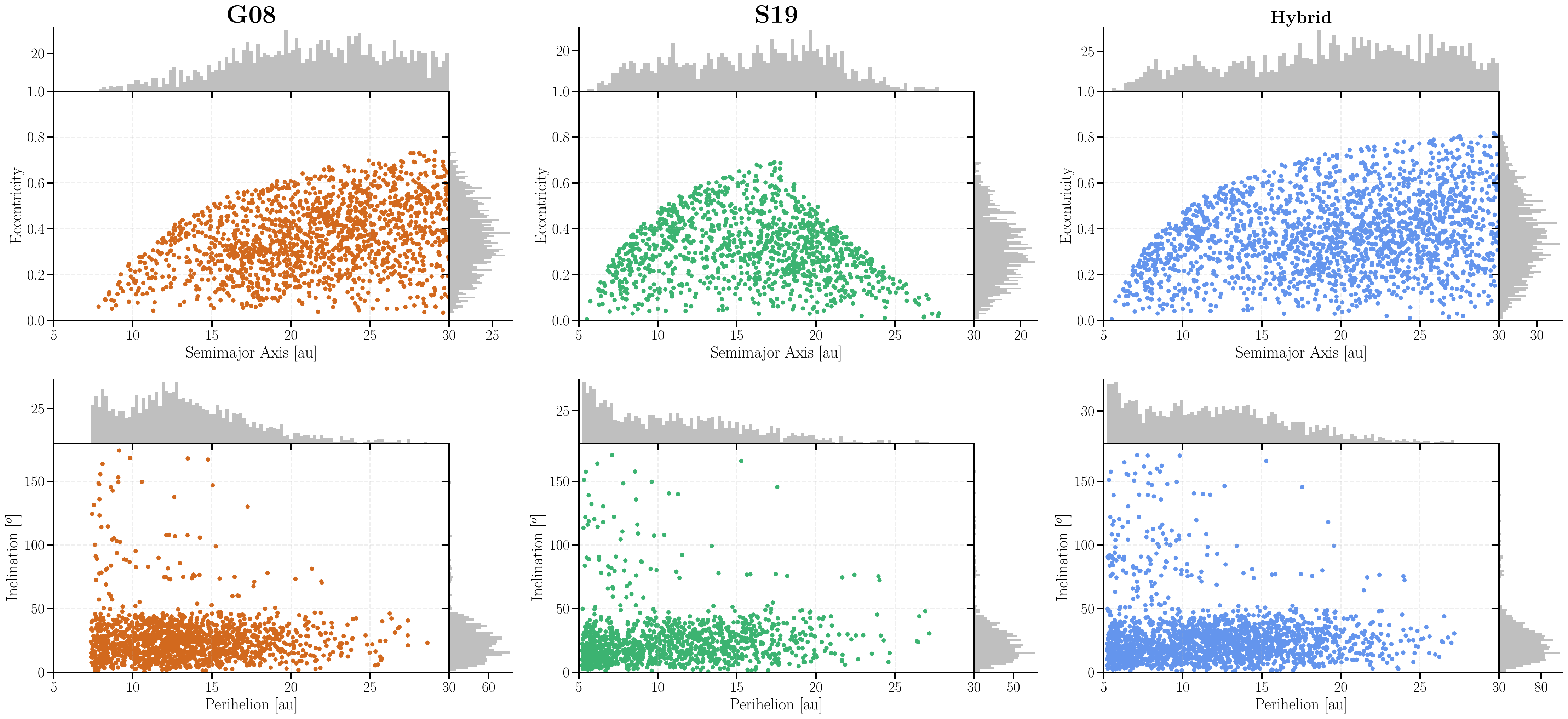}
    \caption{Scatter plots of the orbital spaces covered by discovered objects, similar to Figure \ref{fig:orbitalspace}. Once again the columns represent the three respective definitions, whereas the top row represents Centaur eccentricity as a function of semimajor axis, and the bottom row is Centaur inclination as a function of perihelion distance.}
    \label{fig:discorbs}
\end{figure*}

Figure \ref{fig:10yr} shows the number of unique Centaur discoveries for each definition over the 10 year survey lifetime. We find the total number of Centaurs discovered over the decade to be \update{1524, 1170, and 1967} for the \citetalias{gladman08}, \citetalias{sarid19}, and Hybrid definitions respectively - the total discoveries after years 1, 2, 5, and 10 are highlighted in Table \ref{tab:numbers}, and the orbital space that the discovered Centaurs cover is shown in Figure \ref{fig:discorbs}. These values are based on an assumed color fraction described in Section \ref{sec:2.2.3}, however we have additionally varied this fraction from two extremes of 90:10 to 10:90 blue:red. This has little effect on the yield however, only changing the final \update{number of discovered Centaurs} on the order of \update{$\lesssim10^1$} Centaurs, and so for the rest of this paper we continue analysis using our initial assumption of a 3:1 ratio. \update{The results presented here are the product of one instance of a model simulation, with non-deterministic results due to randomly assigned absolute magnitudes and surface colors, as well as randomization within \texttt{Sorcha} applying survey biases. The uncertainty on 100 simulations of unique models for each definition is $\sim$5-8\%.} 

\begin{deluxetable}{c >{\color{black}} c >{\color{black}} c >{\color{black}} c >{\color{black}} cc}
\tablecaption{Numbers of Centaur discoveries by survey year for each dynamical definition, compared to the number of known Centaurs within the MPC for each corresponding definition. \label{tab:numbers}}
\tablewidth{\columnwidth}
\tablehead{
\colhead{} & \colhead{}     & \colhead{LSST Discovery Numbers} & \colhead{}      & \colhead{}         & \multicolumn{1}{|c}{MPC}  \\
\colhead{} & \colhead{1 yr} & \colhead{2 yrs}                    & \colhead{5 yrs} & \colhead{10 yrs}   & \multicolumn{1}{|c}{Population}
}
\startdata
\citetalias{gladman08} Sample & 767  & 942  & 1240   & 1524  & \multicolumn{1}{|c}{215} \\
\citetalias{sarid19} Sample   & 559  & 682  & 922   & 1170   & \multicolumn{1}{|c}{186} \\   
Hybrid sample                 & 970  & 1195  & 1563  & 1967  & \multicolumn{1}{|c}{288}         
\enddata
\end{deluxetable}

\begin{figure*}
    \centering
    \includegraphics[width=\columnwidth]{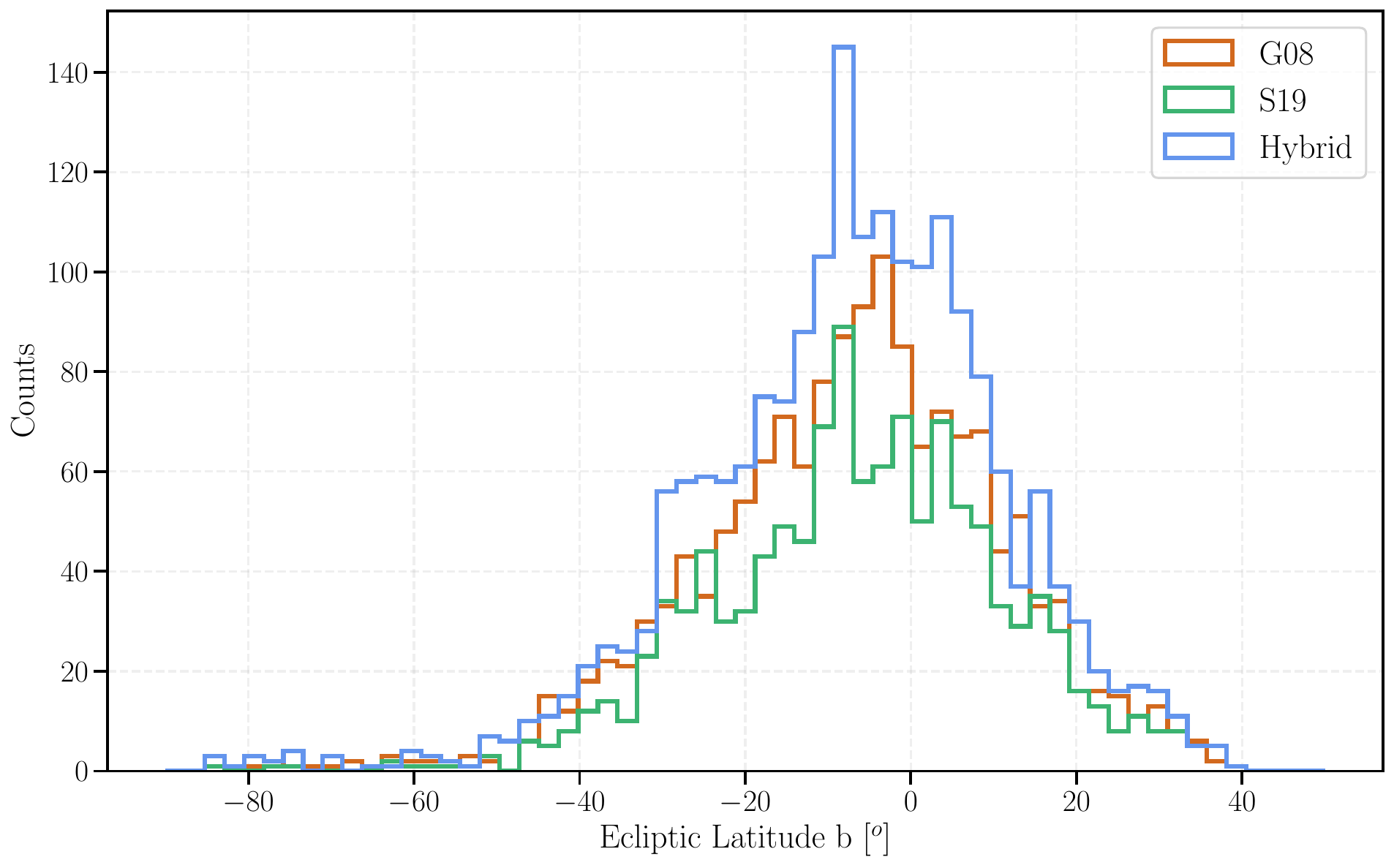}
    \caption{Histograms of the ecliptic latitudes for each of the \citetalias{gladman08}, \citetalias{sarid19}, and Hybrid definitions respectively at the time of each object's discovery.}
    \label{fig:lat}
\end{figure*}

Centaur discovery is seen to be rapid, with 50\% of each population discovered \update{shortly after the beginning} of the second year of \update{the LSST's operation}. Amongst all definitions, year 1's discovery rate stands out compared to all following years. The median \update{r band} apparent magnitude of discovered objects deepens from $\sim$22.7 in year 1 to $\sim$23.5 across the remainder of the survey - the vast majority of bright objects are discovered within the first year, with following years discovery rate being dominated by the LSST limiting magnitude \citep[$\sim$24.7 for \textit{r};][]{ivezic19, bianco22}. Fainter objects are continuously detected as they gain more observations, and so more chances at being detected and passing the SSP linking pipeline. Across all three definitions, \update{$\geq96\%$} are discovered with an ecliptic latitude \update{-40\degree$\leq b \leq40$\degree} as shown in Figure \ref{fig:lat} - however outside of this, \update{between the three models, $\sim$3-14} objects are discovered with ecliptic latitude \update{$\leq-70$\degree}. \update{As most Centaur discoveries have so far been located closer to the ecliptic plane, an increased sample size of high ecliptic latitude Centaurs will offer a unique probe into a more dynamical stable Centaur population, as these objects experience less frequent close encounters with giant planets.}

\begin{figure*}
    \centering
    \includegraphics[width=\columnwidth]{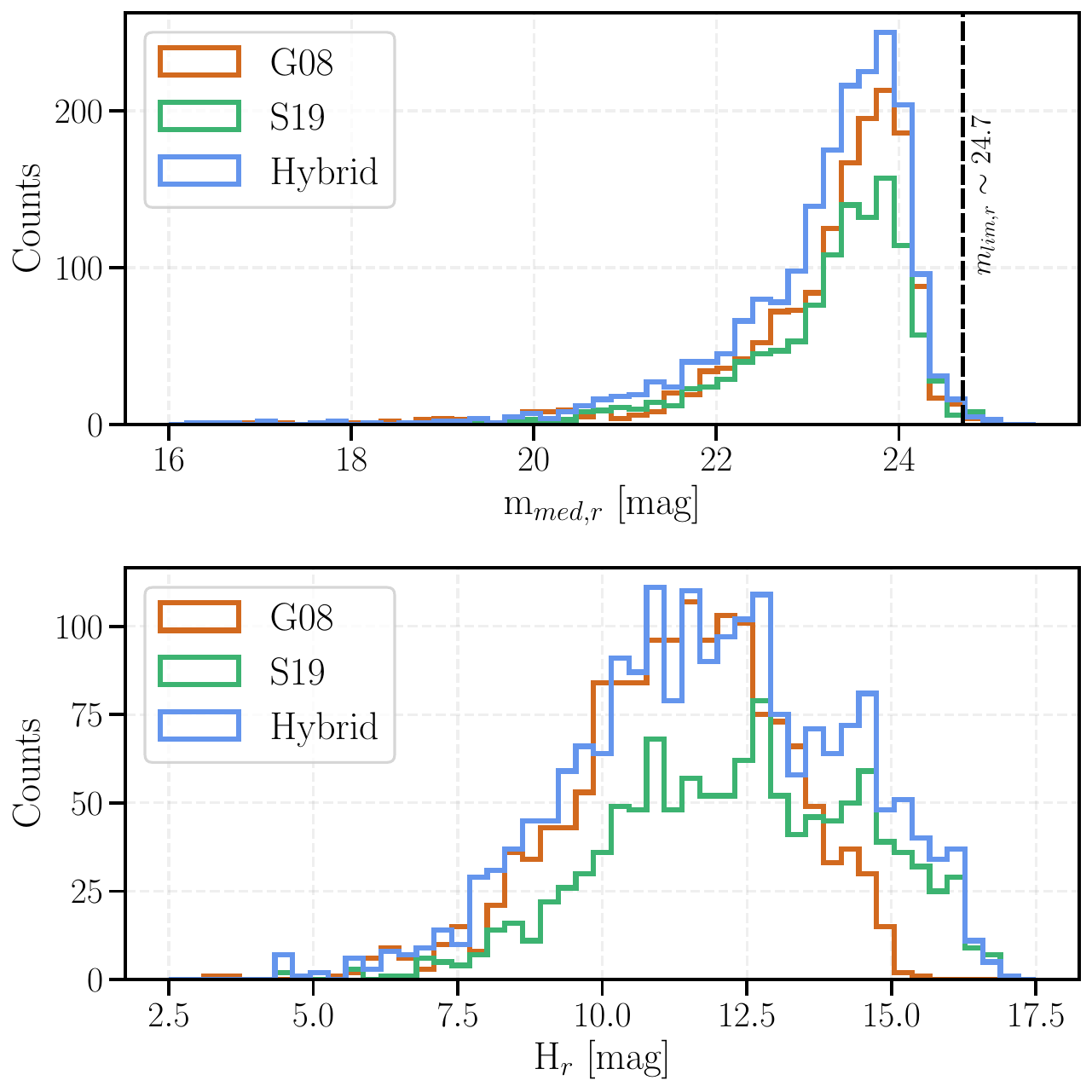}
    \caption{Histogram on top of the median apparent magnitude of discovered Centaurs in the \textit{r}-band, separated by \citetalias{gladman08}, \citetalias{sarid19}, and Hybrid definitions, over-plotted with a dashed line of the limiting magnitude in the LSST \textit{r}. Below is a histogram of the same sample's absolute \textit{r}-band magnitudes.}
    \label{fig:mags}
\end{figure*}

The LSST will discover \update{$\sim$7-12x} more faint objects than are known today, with $\sim$80 Centaurs with \textit{H} $\geq$ 12 recorded in the MPC compared to \update{$\sim$600-950} in all three of our samples. The distributions in the absolute magnitude and median apparent magnitudes of detected Centaurs is shown in Figure \ref{fig:mags}. Table \ref{tab:numbers} also includes a reference to the known MPC numbers for each respective population. Comparing to the 215, 186, and 288 respective \citetalias{gladman08}, \citetalias{sarid19}, and Hybrid definitions, this shows a \update{$\sim$6-7} fold increase in sample size. As a comparison, out of the previous well-characterized Solar System discovery surveys; the Dark Energy Survey discovered a single Centaur \citep[limiting magnitude in \textit{r} band, $m_{lim,r} \sim$ 24.0;][]{bernardinelli22}, OSSOS discovered 20 \citep[$m_{lim,r} \sim$ 24.1 - 25.2;][]{bannister18, cabral19}, Pan-STARRS1 discovered 78 (unknown) \citep[$m_{lim,V} \sim$ 22.5;][]{weryk16}, and the Deep Ecliptic Survey found 13 \citep[$m_{lim,r} \sim$ 26.2;][]{elliot05}. These surveys' primary science goals were not \update{however, }focused on Centaur discovery \update{- they} had differing sky coverage, depths, fields of view, and relative coverage of Centaur orbital space\update{, and their results must be caveated as such}. 

\update{With an increased dataset of Centaurs across a wide orbital space range, the LSST will be able to provide insights into the dynamical transition from Centaur towards JFC.} \updatetwo{\cite{sarid19} posited that existence in the JFC population is preceded by residence in a dynamically short-lived corridor of orbital space known as the `Gateway' region (5.2 au $< a <$ 7.8 au, $e < 0.2$). This region also coincides with observed heliocentric distances which show increases in cometary activity of Centaurs \citep{fraser22, guilbertlepoutre23}.} \update{More recent thermal and dynamical modeling from \cite{guilbertlepoutre23} has however contrasted with this result, showing that only $\sim$20\% of Centaurs are dynamically new to the Gateway prior to becoming JFCs, with several transitions between the two populations frequently occurring, and the majority ($\sim$80\%) transitioning outside of the Gateway altogether. Regardless of dynamical pathway to becoming a JFC, study of objects within this Gateway region offer unique probes into statistically higher processed surfaces of Centaur. Within the \citetalias{sarid19} sample (and \citetalias{sarid19} group within the Hybrid sample), we find $\sim$10-15 objects within the Gateway region - a $\sim$3-5 fold increase on the number of Gateway Centaurs in previous studies \citep{kulyk16, sarid19, steckloff20, hsieh21, kareta21, seligman21, guilbertlepoutre23}. Studies of their surface properties through color analysis (see Section \ref{sec:3.4}) will thus be a means to shed light on the dynamical origins and present-day states of these Centaurs.}


\subsection{Observation Numbers}\label{sec:3.2}
\begin{figure*}
    \centering
    \includegraphics[width=\textwidth]{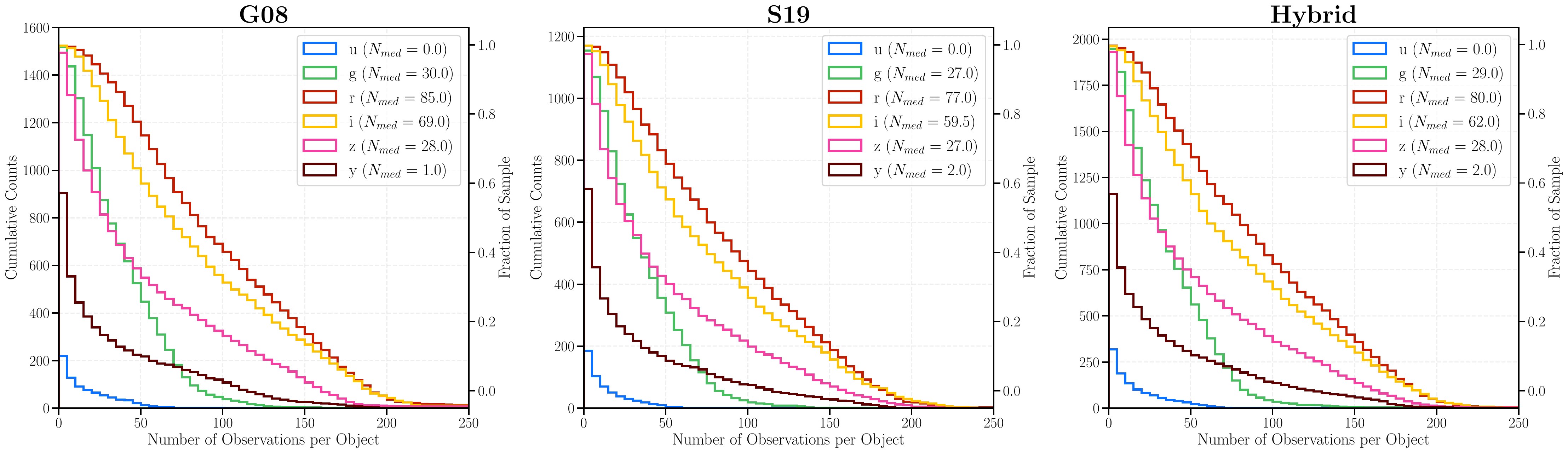}
    \caption{Cumulative histograms of the number of observations each objected detected in each sample will obtain, broken down by filter - excluding those observations that fall within the COSMOS DDF. Note that for all of these plots the x-axis is cut off at 250 for readability, however the \citetalias{gladman08} sample extends to \update{$\sim$800}, the \citetalias{sarid19} sample extends to \update{$\sim$650}, and the Hybrid sample extends to \update{$\sim$900 due to objects entering DDFs}. The reported median values in the legends include these values.}
    \label{fig:nofilt}
\end{figure*}

Over the full 10 years, detected Centaurs will receive a large number of observations, with a median of \update{220, 200, and 207} for the \citetalias{gladman08}, \citetalias{sarid19}, and Hybrid definitions respectively (assuming perfect performance of association and precovery from the SSP \citep{juric20}) - these are further broken down per filter in Figure \ref{fig:nofilt} and  Table \ref{tab:nobs}. The LSST will also continuously observe \update{$\sim60-70\%$} of the discovered objects (or \update{1046, 677, and 1181} discovered \citetalias{gladman08}, \citetalias{sarid19}, and Hybrid Centaurs respectively) as they are observed for approaching the full 10 years of the survey. Per year, this equates to a median of \update{$\sim$24} observations across all filters for all three samples. Observations before linking may only become identified by the SSP pipelines as the orbit is refined. In all three samples we find $\geq$99\% of discovered Centaurs will achieve orbital arcs longer than a year, which will typically enable good orbit determination \citep[see][]{volk24}.  

\begin{deluxetable}{c >{\color{black}} c >{\color{black}} c >{\color{black}} c >{\color{black}} c >{\color{black}} c >{\color{black}} c >{\color{black}} c}
\tablecaption{Breakdown of the median number of observations per object (including zero observations) over the full LSST survey lifetime, as split into each filter for each Centaur definition - excluding observations that fall within the COSMOS DDF. Also shown in the rightmost column is the median number of observations per object across all filters. All columns assume perfect association and precovery from the SSP. \label{tab:nobs}}
\tablewidth{\columnwidth}
\tablehead{
\colhead{}             & \colhead{u} & \colhead{g}                  & \colhead{r}               & \colhead{i}   & \colhead{z} & \colhead{y} & \multicolumn{1}{|c}{All Filters}
}
\startdata
\cutinhead{Median \# of Observations per Object}
\citetalias{gladman08} & 0           & 31                           & 86                        & 69            & 28          & 1           & \multicolumn{1}{|>{\color{black}}c}{220}      \\
\citetalias{sarid19}   & 0           & 28                           & 78                        & 61            & 27          & 2           & \multicolumn{1}{|>{\color{black}}c}{200}      \\
Hybrid                 & 0           & 29                           & 82                        & 63            & 29          & 2           & \multicolumn{1}{|>{\color{black}}c}{207}      \\
\enddata
\end{deluxetable}

Different footprints of the LSST will have different numbers of visits, with DDFs being single fields with a much larger sampling rate than the surrounding WFD area. The number of observations is enhanced for those Centaurs that fall within the DDFs - in particular the COSMOS DDF (centered at 10h 00m 24s / +02d 10m 55s), which at $\sim$9\degree latitude off of the ecliptic makes it particularly sensitive to mildly dynamically excited populations such as the Centaurs. Examining our simulation outputs, \update{30, 35, and 46} Centaurs in the \citetalias{gladman08}, \citetalias{sarid19}, and Hybrid samples (or \update{$\sim$5\%} of each definitions total discoveries), are ever in the COSMOS DDF throughout the 10 year survey. Centaurs will not spend more than one observing season within the 3.5\degree diameter COSMOS DDF - regardless, this is enough time to boost observations to the order of thousands. COSMOS DDF Centaurs receive on average \update{$\sim$400-500} observations, with the median time spent in the COSMOS DDF  being typically \update{$\sim$100} days (however some approach 500 days, and one reaches $\sim$4 years). In comparison, a Centaur that resides \update{solely} in other \update{regions} of the survey footprint will achieve half the number of observations over 10 years. With the sheer number of observations, those few objects that do pass within the COSMOS DDF will make interesting candidates for light-curve and cometary activity analysis.

\subsection{Cometary Activity}\label{sec:3.3}
\updatetwo{Centaur activity, unlike JFCs, is not solely linked to perihelion distance (although most known active Centaurs have a $q <$ 12 au, see \citealt{lilly24}), and has been seen to be possible at any stage of their orbits \citep{peixinho20}. It is thought to be driven by as-of-yet unknown mechanisms rather than the water ice sublimation of JFCs \citep[see][and references therein]{prialnik92, jewitt09, peixinho20}.} The LSST detection estimates outlined in Section \ref{sec:3.1} and Section \ref{sec:3.2}\update{, which are} based on an inactive Centaur population\update{,} thus represent only a lower limit, as activity can cause increases in brightness on the order of $\sim$1-5 magnitudes \citep{kareta20, steckloff20, hsieh21, dobson24}. This in turn will allow for smaller active objects to be observable within our samples, particularly in the innermost regions of Centaur orbits where activity is mostly observed. Estimates from \cite{jewitt09, peixinho20, chandler24} all place the activity occurrence rate for Centaurs to be $\sim$9-13\%, which represents in our samples \update{$\sim$100-250} active Centaurs observed.

Recent work by \cite{lilly24} has shown evidence of jumps in semimajor axis over months-years being a precursor to the onset of such cometary activity, with the need for monitoring to probe this potential trend. \cite{mazzotta18} also \update{discuss} the need for the monitoring of already active Centaurs and their surface colors as a means of following the evolution of the previously discussed bimodal color distribution. With observational arcs approaching a decade for hundreds of Centaurs, the LSST is able to provide just such a monitoring service for these objects. With \update{$\sim$700-1100} discovered Centaurs each gaining \update{$\sim$24} observations a year, activity searches will be possible in the form of probing extensions of the PSF of each object or direct coma detection \citep[e.g.][]{jewitt09, mazzotta18, seccull19, dobson24}. Alongside this, deviations in the phase curves of each object will also be usable as a means to search for activity \citep[e.g.][]{dobson23, dobson24} - see Section \ref{sec:3.4} for further discussion \update{on the predictions for Centaur phase curve measurements within the LSST}. 

\subsection{Colors, Light-curves, and Phase Curves}\label{sec:3.4}
Previous surveys have been predominantly in single filters - with 6 filters available the LSST will be able to naturally perform surface color, light-curve, and phase curve studies without the need for follow-up. Across all samples there are \updatetwo{fewer} observations in the \textit{u} and \textit{y} bands compared to \textit{griz} (see Table \ref{tab:nobs}). The \textit{u} band suffers a result of survey cadence and object colors being fainter in this wavelength range, whereas \textit{y} band measurements are hindered by survey cadence and near-infrared sky background lowering the limiting magnitude of these observations \citep{bianco22}. For all three samples the median signal-to-noise ratio (S/N) for \textit{griz} measurements is \update{$\sim$8} (or a photometric precision of \update{$\sim$0.125} mag). With \update{$\sim$24} observations per year, the Centaur population will be provided ample opportunities for characterization. In order to probe the quantity and quality of surface colors, light-curves, and phase curves that are potentially available within our samples, we look to \cite{schwamb23} and their set of metrics. These metrics, based on LSST Metrics Analysis Frameworks (MAF) within \package{rubin\_sim} \citep{jones14}, assume at least 30 observations in one filter and 20 in a second filter per object, all at a S/N $\geq$ 5. Whilst not a guarantee for precise measurement, these metrics provide a first approximation to determining sparse light curves, as well as fitting phase curves, which will allow for the determining surface colors.

\begin{figure*}
    \centering
    \includegraphics[scale=0.4]{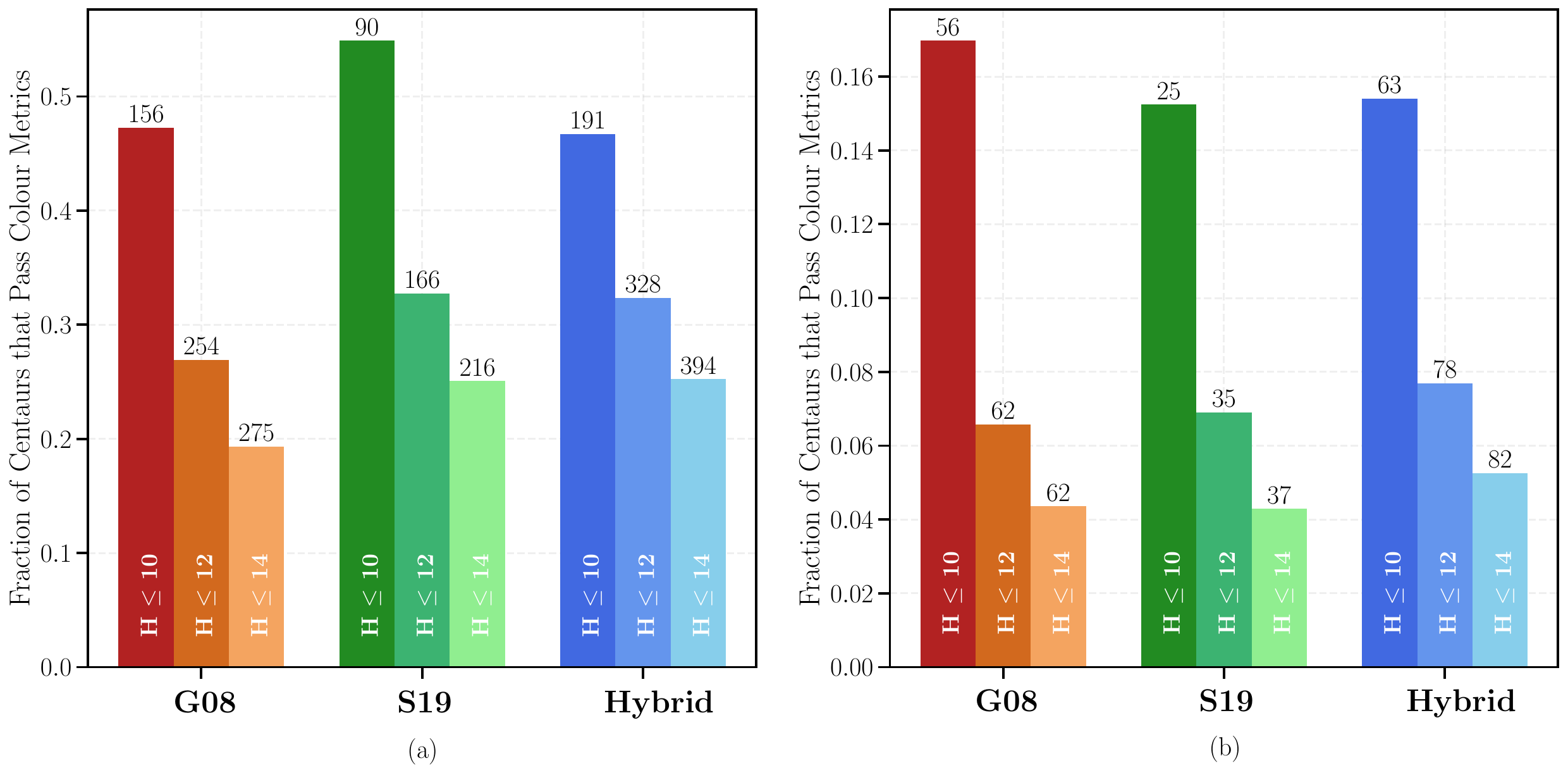}
    \caption{Number of Centaurs which would obtain at least three colors in \textit{ugrizy}. Each bar represents the fraction of each sub-sample of discovered objects that would pass the color metrics defined in \cite{schwamb23}, with the actual number of Centaurs this represents above each bar. (a) highlights those passing the metrics with a SNR$\geq$5 cut, whereas (b) highlights the same but for a SNR$\geq$25 cut as was used in the OSSOS survey.}
    \label{fig:colmetrics}
\end{figure*}

We apply these metrics in Figure \ref{fig:colmetrics}, with \update{$\sim$210-400} $H \leq 14$ (our sample size drops fainter than this absolute magnitude for all three definitions, see Figure \ref{fig:mags}) Centaurs across the three samples obtaining three colors across \textit{ugrizy}. This is \update{at least} an order of magnitude increase from the samples used to investigate bimodality in the \cite{tegler98, tegler03} (\textit{N} = 5, 3), \cite{peixinho03} (\textit{N} = 18), \cite{wong16} (\textit{N} = 15), and a \update{$\sim$4 fold increase on} the \cite{tegler16} studies (\textit{N} = 50). Alternative color studies, such as the Colours of the Outer Solar System Origins Survey (Col-OSSOS) \citep{schwamb19, fraser23} use a differing threshold of S/N $\geq$ 25 in order to create a more precisely measured sample of objects in order to investigate ice line transitions in the protoplanetary disk. We additionally modify the \cite{schwamb23} metrics to increase the S/N requirement to create a Col-OSSOS quality sub-sample, shown in Figure \ref{fig:colmetrics}. With \update{$\sim$40-80} $H \leq 14$ Centaurs across the three samples, this massively increases the single Centaur sample size within Col-OSSOS. With such an expanded sample, the origin of Centaurs in the primordial disk will be further able to be probed, as well as investigating the effects of thermal processing on surface colors. Regardless of which metric used, both results here highlight the LSST's ability to create expanded studies into Centaur evolutionary history, with the potential for ultra precisely measured surface colors.

Phase curves remain an important method for probing an object's overall surface properties over changing viewing angles. For all three definitions of our objects, we sample a phase angle range less than $\sim$9-14\degree, with the median phase angle at time of discovery being $\sim$2\degree. The prior metrics will allow for determination of quality phase curves as a by-product of the criteria required for light-curve measurement. We also look at phase curve-specific quality metrics for our sample in order to determine the density and quality that will be available naturally through the LSST cadence. We fit observations with a linear slope with \package{curve\_fit} from \package{scipy}, using parameters determined from a Monte Carlo synthetic phase curve algorithm, trialing 10,000 phase curves within the Gaussian uncertainties of each data point. After fitting, we make a cut on phase curve quality using a modified set of metrics as initially defined in \cite{robinson24} for Jupiter Trojans, a set of objects also observed in a wide-field survey, albeit with slightly larger phase angle range behavior. The cuts as listed below allow for removal of observations with poor photometric precision, as well as limiting spread in the dataset by placing limits on the phase curve fit itself. The cuts used are as follows:

\begin{enumerate}[label=(\alph*)]
    \item The number of photometric points $\geq$ 25
    \item The phase angle range $\alpha_{max}$ - $\alpha_{min}$ $\geq$ 3\degree
    \item The absolute magnitude uncertainty $\sigma_H$ $\leq$ 0.1 mag
    \item The best fit linear phase coefficient uncertainty $\sigma_\beta$ $\leq$ 0.02 mag deg$^{-1}$
\end{enumerate}

\begin{figure*}
    \includegraphics[width=\textwidth]{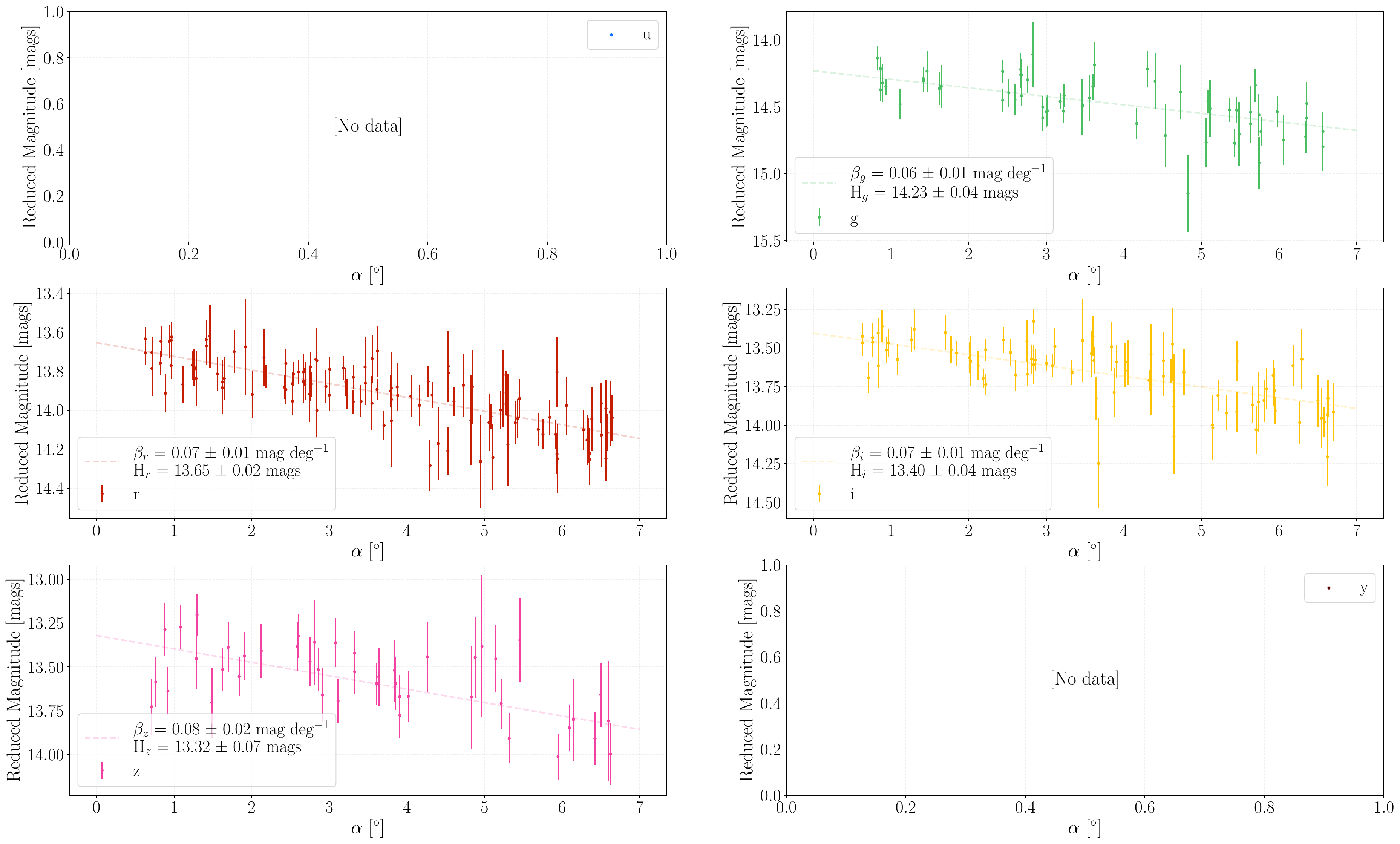}\centering
    \caption{Example phase curves across all \textit{ugrizy} filters for a randomly sampled object from the \citetalias{gladman08} sample. There are no \textit{uy} band phase curves here as they do not have enough observations passing metrics for quality phase curves. All filters that do pass the metrics have been overplotted with the linear fit to the data as determined in the text.}
    \label{fig:phasecurves}
\end{figure*}

The overall numbers of phase curves available for each definition highlighted in Table \ref{tab:phasecurves}, with an example \citetalias{gladman08} object's \textit{ugrizy} phase curves shown in Figure \ref{fig:phasecurves}. \update{Here, we show} an order of magnitude increase on the 23 Centaurs used in the \cite{ayalaloera18}, \cite{alvarezcandal16}, and \cite{rabinowitz07} studies. Comparing the $H_r$ uncertainties to those of the $H_V$ uncertainties reported by the JPL Small Body Database\footnote{\href{https://ssd.jpl.nasa.gov/tools/sbdb_query.html}{https://ssd.jpl.nasa.gov/tools/sbdb\_query.html}} , only $\sim$80 values have uncertainties, with a median uncertainty of 0.33 mag - \update{our sample of detected objects have median measured uncertainties an order of magnitude lower than this at $\sim$0.03 mag in the \textit{griz} filters.} With \update{$\sim$9-14} fold more absolute magnitude measurements available from \textit{r}-band phase curves alone, the LSST will be capable of massively increasing the sample size of absolute magnitudes of Centaurs currently available. 

\begin{deluxetable*}{c >{\color{black}} c >{\color{black}} c >{\color{black}} c >{\color{black}} c >{\color{black}} c >{\color{black}} c}[h!]
\tabletypesize{\scriptsize}
\tablewidth{\columnwidth}
\centering\tablecaption{Number of high quality phase curves available per filter for each sample definition after applying the metrics described in Section \ref{sec:3.4}. \label{tab:phasecurves}}
\tablewidth{\columnwidth}
\tablehead{
\colhead{} & \colhead{\# \textit{u}} & \colhead{\# \textit{g}} & \colhead{\# \textit{r}}  & \colhead{\# \textit{i}} & \colhead{\# \textit{z}} & \colhead{\# \textit{y}}
}
\startdata
\citetalias{gladman08} Sample & 34   & 456  & 825  & 755  & 451  & 162    \\
\citetalias{sarid19} Sample   & 12   & 411  & 736  & 663  & 421  & 154    \\   
Hybrid sample                 & 29   & 629  & 1112 & 990  & 622  & 238                     
\enddata
\end{deluxetable*}

\section{Conclusions}\label{sec:4}

\updatetwo{In this work, we have used the current best dynamical model that describes the Centaur population \citep{nesvorny19}, combined with real, accurate colors and phase curve parameters in order to investigate the Centaur yield within the LSST. We have also showcased the potential for characterizing them through light and phase curve analysis.} In order to do this we used the survey simulator \texttt{Sorcha} \citep{merritt25, holman25} combined with the current latest survey cadence simulations \citep{scocv3} in order to completely forward bias the output simulated detections. We investigated this for three different subset definitions of Centaurs in order to investigate the impact this will have on discoverability and characterization. Our main conclusions are summarized as follows:

\begin{itemize}
    \item Across all three definitions, the predicted yield of Centaur discoveries is set to increase the known \update{MPC population of 215, 186, and 288 Centaurs} by \update{$\sim$7-12x}, giving \update{1524, 1170, and 1967} discoveries for the \citetalias{gladman08}, \citetalias{sarid19}, and Hybrid samples respectively. The lower yields for \citetalias{gladman08} and \citetalias{sarid19} represent a more stringent dynamical definitional criteria rather than a decrease in discovered objects.
    \item The precise color fraction of blue:red objects within the population does not significantly affect this yield nor the rate of discovery of Centaurs, even at either extreme ends of simulation\update{, varying on the order of $\lesssim10^1$ discovered Centaurs}. 
    \item Discovery is predicted to happen relatively early - \update{50\%} population completion is expected to occur within \update{2 years of survey operation}. The majority of (on average) brighter objects (i.e. $m \leq$23$^{rd}$ mag) are discovered within the first year alone, with the remaining years' discoveries being driven by a combination of fainter objects at the limiting magnitudes gaining enough observations to pass linking metrics.
    \item The DDF COSMOS presents an exciting and rich opportunity for Centaur characterization. Only \update{$\sim$30-50} observable Centaurs enter the field, and by spending on average merely 2 months within the DDF, they will receive twice as many observations as that of an object that never enters will have. At $\sim$500 photometric points, this represents an important means of constructing high quality, dense phase and light curves for Centaurs for detailed analysis.
    \item The median observational arc for the Centaurs approaches the full 10 year baseline, with a median $\sim$200 observations per object - or $\sim$20 per object per year across the \textit{ugrizy} range. This in and of itself represents a key opportunity for discovery of ongoing cometary-like activity by searches for PSF extension or direct coma detection, as well as its monitoring by way of light and phase curve analysis.
    \item Applying color metrics based on the number of and measured S/N of observations, we see upwards of \update{$\sim$200-300} Centaurs with at least three \textit{ugrizy} color measurements to a photometric precision of 0.2 mag, the typical catalog color uncertainty of objects in literature. Constraining to a larger S/N yields an ultra-well defined sample of up to \update{$\sim$40-80} objects.
    \item By applying a set of cuts on the data to ensure quality phase curves, \update{$\sim$300-500} phase curves are available for investigation in the \textit{griz} filters. Fitting these datasets with linear models shows that median uncertainties in the absolute magnitudes determined here are on the order of $\sim$0.03 mag, almost doubling the current sample of known absolute magnitudes, and bringing the uncertainty on these values down by an order of magnitude of what is currently known in the MPC.
\end{itemize}

\updatetwo{Our predictions for the Centaur yield and characterization within the LSST are dependent on the size of the input model population. We have used a model population using the scaling constants from \cite{kurlander24} as this represents the best constrained value in the literature and provides the most realistic model. However, we note that utilizing alternative scaling changes the expected LSST Centaur yield. Employing the OSSOS observationally matched scaling of $N_0$ = 21,000 Centaurs with $H_r < 13.7$ from \cite{nesvorny19} sees a decrease in population size of $\sim$2\% (or, $\sim$177k objects) for the \citetalias{gladman08} model. This in turn results in an identical $\sim$2\% drop in detections (or, $\sim$30 objects) and other associated light curve and surface color metrics. Employing the alternative scaling derived from Jupiter Trojan size distributions from \cite{nesvorny19} of $N_0$ = 15,600 Centaurs with $H_r < 13.7$ results in an even larger drop of $\sim$33\% (or, 2.57$\times10^6$ objects), resulting in $\sim400$ fewer detected objects for the \citetalias{gladman08} model. The same results are true for the alternative Centaur definitions. Despite this, the maximum variation in discovered objects is only $\sim$400 objects between the different $H$ distributions. \updatethree{The same result is true of our choice of single power law for the $H$ distribution - with only a $<1\%$ effect on the overall detected numbers, this is drowned out by the random variation of $\sim$5-8\% between runs. With less low-$H$ objects, there will be less opportunities for detailed light curve, phase curve, and surface color studies, but only on the order of 10's of objects.} The choice of $N_0$ scaling\updatethree{, or functional form of $H$ distribution,} does not however change our overall conclusions. The LSST will still be transformative in the discovery and characterization of Centaurs, increasing on the known $\sim$300 MPC sample.}

This analysis \updatetwo{also} has the caveat of depending on perfect template generation within the LSST for difference imaging. Regardless of year 1 template generation status, observations will be able to be recovered as the first-year Data Release is shared $\sim$1 year after operation begins. \cite{schwamb21} and \cite{robinson24b} discuss the impact that year 1 template generation has on the observability of small objects, with \cite{schwamb23} investigating several scenarios of observational cadence. Comparing our predictions for the one snap realization of the v4.0 cadence to the two snap, we see very little variation in the overall trend or numbers of detections and their potential for characterization.

The immense potential in the long baseline, depth, and high cadence of the LSST for Centaur discoverability and characterization is readily apparent in the rapid rate of Centaur discovery, the dense number of observations they acquire, and the photometric precisions that are able to be reached for color and phase curve measurements. The LSST represents a transformative opportunity to understand the Centaur population's evolution and dynamics through the data gathered within its decade-long operation. Early science, that is any science possible through and including the first-year data release, especially represents a rich opportunity for future research, including additional follow-up observations for any objects discovered within the first year with supporting ground-based facilities. Due to the quantity and quality of this data, model testing and refinement with survey simulators like \texttt{Sorcha} will also be readily available within the first months of operation of the LSST \updatethree{as progressively debiased LSST datasets become available.}


%

\section*{Acknowledgments}

 This work was supported by a LSST Discovery Alliance LINCC Frameworks Incubator grant [2023-SFF-LFI-01-Schwamb]. Support was provided by Schmidt Sciences. J.M. acknowledges support from the Department for the Economy (DfE) Northern Ireland postgraduate studentship scheme and travel support from the STFC for UK participation in LSST through grant ST/S006206/1. M.E.S. and S.R.M. acknowledge support in part from UK Science and Technology Facilities Council (STFC) grants ST/V000691/1 and ST/X001253/1. M.J. and P.H.B. and J.A.K. acknowledge the support from the University of Washington College of Arts and Sciences, Department of Astronomy, and the DiRAC Institute. The DiRAC Institute is supported through generous gifts from the Charles and Lisa Simonyi Fund for Arts and Sciences and the Washington Research Foundation. M.J. wishes to acknowledge the support of the Washington Research Foundation Data Science Term Chair fund, and the University of Washington Provost's Initiative in Data-Intensive Discovery. J.M and J.A.K. thank the LSST-DA Data Science Fellowship Program, which is funded by LSST-DA, the Brinson Foundation, and the Moore Foundation; their participation in the program has benefited this work. K.V. acknowledges support from NASA (grants 80NSSC23K1169 and 80NSSC23K0886). S.C. and S.E. acknowledge support by the National Science Foundation through Award AST-2307570. Any opinions, findings, and conclusions or recommendations expressed in this material are those of the authors and do not necessarily reflect the views of the National Science Foundation.
 
 This work was also supported via the Preparing for Astrophysics with LSST Program, funded by the Heising Simons Foundation through grant 2021-2975, and administered by Las Cumbres Observatory. This work was supported in part by the LSST Discovery Alliance Enabling Science grants program, the B612 Foundation, the University of Washington's DiRAC (Data-intensive Research in Astrophysics and Cosmology) Institute, the Planetary Society, Karman+, and  Breakthrough Listen, Adler Planetarium through generous support of the LSST Solar System Readiness Sprints. Breakthrough Listen is managed by the Breakthrough Initiatives, sponsored by the Breakthrough Prize Foundation (\url{http://www.breakthroughinitiatives.org}).

 This research has made use of NASA’s Astrophysics Data System Bibliographic Services. This research has made use of data and/or services provided by the International Astronomical Union's Minor Planet Center. The SPICE Resource files used in this work are described in \citep{acton18, acton96}. Simulations in this paper made use of the REBOUND N-body code \citep{rein12}. The simulations were integrated using IAS15, a 15th order Gauss-Radau integrator \citep{rein15}. Some of the results in this paper have been derived using the healpy and HEALPix packages. This work made use of Astropy:\footnote{\href{http://www.astropy.org}{http://www.astropy.org}} a community-developed core Python package and an ecosystem of tools and resources for astronomy \citep{astropy13, astropy18, astropy22}.

 This material or work is supported in part by the National Science  Foundation through Cooperative Agreement AST-1258333 and Cooperative Support Agreement AST1836783 managed by the Association of Universities for Research in Astronomy (AURA), and the Department of Energy under Contract No. DE-AC02-76SF00515 with the SLAC National Accelerator Laboratory managed by Stanford University.  

We are grateful for use of the computing resources from the Northern Ireland High Performance Computing (NI-HPC) service funded by EPSRC (EP/T022175). We gratefully acknowledge the support of the Center for Advanced Computing and Modelling, University of Rijeka (Croatia), for providing supercomputing resources at HPC (High Performance Computing) Bura.

We are particularly grateful to Noem\'{i} Pinilla-Alonso for providing the original data for the spectra of (5145) Pholus and (54598) Bienor, from which the color analysis of this work was based on. \update{We also thank the anonymous referee for their constructive feedback which improved this manuscript.}

The authors wish to acknowledge the researchers who worked tirelessly to rapidly develop COVID-19 vaccines and subsequent boosters. Without all their efforts, we would not have been able to pursue this work. 

Data Access:  The software simulator used in this work is available open-source at \url{https://github.com/dirac-institute/Sorcha}. The rubin$\_$sim LSST cadence simulation databases are available at \url{https://s3df.slac.stanford.edu/data/rubin/sim-data/}.


\software{Sorcha \citep{merritt25, holman25}, ASSIST \citep{holman23, rein23}, Astropy \citep{astropy13, astropy18}, Healpy \citep{zonca19, gorski05}, Matplotlib \citep{hunter07}, Numba \citep{lam15}, Numpy \citep{harris20}, Pandas \citep{mckinney10, pandas20}, Pooch \citep{uieda20}, PyTables \citep{pytables02}, REBOUND \citep{rein12, rein15}, rubin\_sim \citep{bianco22, yoachim23}, rubin\_scheduler \citep{naghib19, yoachim24b}, sbpy \citep{mommert19}, SciPy \citep{virtanen20}, Spiceypy \citep{annex20}, sqlite (\href{https://www.sqlite.org/index.html}{https://www.sqlite.org/index.html}), sqlite3 (\href{https://docs.python.org/3/library/sqlite3.html}{https://docs.python.org/3/library/sqlite3.html}), tqdm \citep{dacostaluis23}, Jupyter Notebook \citep{kluyver16}, seaborn \citep{waskom21}, CMasher \citep{vandervelden20}}






\bibliography{refs}{}

\begin{thebibliography}{}
\expandafter\ifx\csname natexlab\endcsname\relax\def\natexlab#1{#1}\fi
\providecommand{\url}[1]{\href{#1}{#1}}
\providecommand{\dodoi}[1]{doi:~\href{http://doi.org/#1}{\nolinkurl{#1}}}
\providecommand{\doeprint}[1]{\href{http://ascl.net/#1}{\nolinkurl{http://ascl.net/#1}}}
\providecommand{\doarXiv}[1]{\href{https://arxiv.org/abs/#1}{\nolinkurl{https://arxiv.org/abs/#1}}}

\bibitem[{{Acton} {et~al.}(2018){Acton}, {Bachman}, {Semenov}, \& {Wright}}]{acton18}
{Acton}, C., {Bachman}, N., {Semenov}, B., \& {Wright}, E. 2018, \planss, 150, 9, \dodoi{10.1016/j.pss.2017.02.013}

\bibitem[{{Acton}(1996)}]{acton96}
{Acton}, C.~H. 1996, \planss, 44, 65, \dodoi{10.1016/0032-0633(95)00107-7}

\bibitem[{{Adams} {et~al.}(2014){Adams}, {Gulbis}, {Elliot}, {Benecchi}, {Buie}, {Trilling}, \& {Wasserman}}]{adams14}
{Adams}, E.~R., {Gulbis}, A.~A.~S., {Elliot}, J.~L., {et~al.} 2014, \aj, 148, 55, \dodoi{10.1088/0004-6256/148/3/55}

\bibitem[{{Alvarez-Candal} {et~al.}(2019){Alvarez-Candal}, {Ayala-Loera}, {Gil-Hutton}, {Ortiz}, {Santos-Sanz}, \& {Duffard}}]{alvarezcandal19}
{Alvarez-Candal}, A., {Ayala-Loera}, C., {Gil-Hutton}, R., {et~al.} 2019, \mnras, 488, 3035, \dodoi{10.1093/mnras/stz1880}

\bibitem[{{Alvarez-Candal} {et~al.}(2008){Alvarez-Candal}, {Fornasier}, {Barucci}, {de Bergh}, \& {Merlin}}]{alvarezcandal08}
{Alvarez-Candal}, A., {Fornasier}, S., {Barucci}, M.~A., {de Bergh}, C., \& {Merlin}, F. 2008, \aap, 487, 741, \dodoi{10.1051/0004-6361:200809705}

\bibitem[{{Alvarez-Candal} {et~al.}(2016){Alvarez-Candal}, {Pinilla-Alonso}, {Ortiz}, {Duffard}, {Morales}, {Santos-Sanz}, {Thirouin}, \& {Silva}}]{alvarezcandal16}
{Alvarez-Candal}, A., {Pinilla-Alonso}, N., {Ortiz}, J.~L., {et~al.} 2016, \aap, 586, A155, \dodoi{10.1051/0004-6361/201527161}

\bibitem[{{Annex} {et~al.}(2020){Annex}, {Pearson}, {Seignovert}, {Carcich}, {Eichhorn}, {Mapel}, {von Forstner}, {McAuliffe}, {del Rio}, {Berry}, {Aye}, {Stefko}, {de Val-Borro}, {Kulumani}, \& {Murakami}}]{annex20}
{Annex}, A., {Pearson}, B., {Seignovert}, B., {et~al.} 2020, The Journal of Open Source Software, 5, 2050, \dodoi{10.21105/joss.02050}

\bibitem[{{Annis} {et~al.}(2014){Annis}, {Soares-Santos}, {Strauss}, {Becker}, {Dodelson}, {Fan}, {Gunn}, {Hao}, {Ivezi{\'c}}, {Jester}, {Jiang}, {Johnston}, {Kubo}, {Lampeitl}, {Lin}, {Lupton}, {Miknaitis}, {Seo}, {Simet}, \& {Yanny}}]{annis14}
{Annis}, J., {Soares-Santos}, M., {Strauss}, M.~A., {et~al.} 2014, \apj, 794, 120, \dodoi{10.1088/0004-637X/794/2/120}

\bibitem[{{Astropy Collaboration} {et~al.}(2013){Astropy Collaboration}, {Robitaille}, {Tollerud}, {Greenfield}, {Droettboom}, {Bray}, {Aldcroft}, {Davis}, {Ginsburg}, {Price-Whelan}, {Kerzendorf}, {Conley}, {Crighton}, {Barbary}, {Muna}, {Ferguson}, {Grollier}, {Parikh}, {Nair}, {Unther}, {Deil}, {Woillez}, {Conseil}, {Kramer}, {Turner}, {Singer}, {Fox}, {Weaver}, {Zabalza}, {Edwards}, {Azalee Bostroem}, {Burke}, {Casey}, {Crawford}, {Dencheva}, {Ely}, {Jenness}, {Labrie}, {Lim}, {Pierfederici}, {Pontzen}, {Ptak}, {Refsdal}, {Servillat}, \& {Streicher}}]{astropy13}
{Astropy Collaboration}, {Robitaille}, T.~P., {Tollerud}, E.~J., {et~al.} 2013, \aap, 558, A33, \dodoi{10.1051/0004-6361/201322068}

\bibitem[{{Astropy Collaboration} {et~al.}(2018){Astropy Collaboration}, {Price-Whelan}, {Sip{\H{o}}cz}, {G{\"u}nther}, {Lim}, {Crawford}, {Conseil}, {Shupe}, {Craig}, {Dencheva}, {Ginsburg}, {VanderPlas}, {Bradley}, {P{\'e}rez-Su{\'a}rez}, {de Val-Borro}, {Aldcroft}, {Cruz}, {Robitaille}, {Tollerud}, {Ardelean}, {Babej}, {Bach}, {Bachetti}, {Bakanov}, {Bamford}, {Barentsen}, {Barmby}, {Baumbach}, {Berry}, {Biscani}, {Boquien}, {Bostroem}, {Bouma}, {Brammer}, {Bray}, {Breytenbach}, {Buddelmeijer}, {Burke}, {Calderone}, {Cano Rodr{\'\i}guez}, {Cara}, {Cardoso}, {Cheedella}, {Copin}, {Corrales}, {Crichton}, {D'Avella}, {Deil}, {Depagne}, {Dietrich}, {Donath}, {Droettboom}, {Earl}, {Erben}, {Fabbro}, {Ferreira}, {Finethy}, {Fox}, {Garrison}, {Gibbons}, {Goldstein}, {Gommers}, {Greco}, {Greenfield}, {Groener}, {Grollier}, {Hagen}, {Hirst}, {Homeier}, {Horton}, {Hosseinzadeh}, {Hu}, {Hunkeler}, {Ivezi{\'c}}, {Jain}, {Jenness}, {Kanarek}, {Kendrew}, {Kern}, {Kerzendorf}, {Khvalko}, {King}, {Kirkby}, {Kulkarni},
  {Kumar}, {Lee}, {Lenz}, {Littlefair}, {Ma}, {Macleod}, {Mastropietro}, {McCully}, {Montagnac}, {Morris}, {Mueller}, {Mumford}, {Muna}, {Murphy}, {Nelson}, {Nguyen}, {Ninan}, {N{\"o}the}, {Ogaz}, {Oh}, {Parejko}, {Parley}, {Pascual}, {Patil}, {Patil}, {Plunkett}, {Prochaska}, {Rastogi}, {Reddy Janga}, {Sabater}, {Sakurikar}, {Seifert}, {Sherbert}, {Sherwood-Taylor}, {Shih}, {Sick}, {Silbiger}, {Singanamalla}, {Singer}, {Sladen}, {Sooley}, {Sornarajah}, {Streicher}, {Teuben}, {Thomas}, {Tremblay}, {Turner}, {Terr{\'o}n}, {van Kerkwijk}, {de la Vega}, {Watkins}, {Weaver}, {Whitmore}, {Woillez}, {Zabalza}, \& {Astropy Contributors}}]{astropy18}
{Astropy Collaboration}, {Price-Whelan}, A.~M., {Sip{\H{o}}cz}, B.~M., {et~al.} 2018, \aj, 156, 123, \dodoi{10.3847/1538-3881/aabc4f}

\bibitem[{{Astropy Collaboration} {et~al.}(2022){Astropy Collaboration}, {Price-Whelan}, {Lim}, {Earl}, {Starkman}, {Bradley}, {Shupe}, {Patil}, {Corrales}, {Brasseur}, {N{\"o}the}, {Donath}, {Tollerud}, {Morris}, {Ginsburg}, {Vaher}, {Weaver}, {Tocknell}, {Jamieson}, {van Kerkwijk}, {Robitaille}, {Merry}, {Bachetti}, {G{\"u}nther}, {Aldcroft}, {Alvarado-Montes}, {Archibald}, {B{\'o}di}, {Bapat}, {Barentsen}, {Baz{\'a}n}, {Biswas}, {Boquien}, {Burke}, {Cara}, {Cara}, {Conroy}, {Conseil}, {Craig}, {Cross}, {Cruz}, {D'Eugenio}, {Dencheva}, {Devillepoix}, {Dietrich}, {Eigenbrot}, {Erben}, {Ferreira}, {Foreman-Mackey}, {Fox}, {Freij}, {Garg}, {Geda}, {Glattly}, {Gondhalekar}, {Gordon}, {Grant}, {Greenfield}, {Groener}, {Guest}, {Gurovich}, {Handberg}, {Hart}, {Hatfield-Dodds}, {Homeier}, {Hosseinzadeh}, {Jenness}, {Jones}, {Joseph}, {Kalmbach}, {Karamehmetoglu}, {Ka{\l}uszy{\'n}ski}, {Kelley}, {Kern}, {Kerzendorf}, {Koch}, {Kulumani}, {Lee}, {Ly}, {Ma}, {MacBride}, {Maljaars}, {Muna}, {Murphy}, {Norman},
  {O'Steen}, {Oman}, {Pacifici}, {Pascual}, {Pascual-Granado}, {Patil}, {Perren}, {Pickering}, {Rastogi}, {Roulston}, {Ryan}, {Rykoff}, {Sabater}, {Sakurikar}, {Salgado}, {Sanghi}, {Saunders}, {Savchenko}, {Schwardt}, {Seifert-Eckert}, {Shih}, {Jain}, {Shukla}, {Sick}, {Simpson}, {Singanamalla}, {Singer}, {Singhal}, {Sinha}, {Sip{\H{o}}cz}, {Spitler}, {Stansby}, {Streicher}, {{\v{S}}umak}, {Swinbank}, {Taranu}, {Tewary}, {Tremblay}, {de Val-Borro}, {Van Kooten}, {Vasovi{\'c}}, {Verma}, {de Miranda Cardoso}, {Williams}, {Wilson}, {Winkel}, {Wood-Vasey}, {Xue}, {Yoachim}, {Zhang}, {Zonca}, \& {Astropy Project Contributors}}]{astropy22}
{Astropy Collaboration}, {Price-Whelan}, A.~M., {Lim}, P.~L., {et~al.} 2022, \apj, 935, 167, \dodoi{10.3847/1538-4357/ac7c74}

\bibitem[{{Ayala-Loera} {et~al.}(2018){Ayala-Loera}, {Alvarez-Candal}, {Ortiz}, {Duffard}, {Fern{\'a}ndez-Valenzuela}, {Santos-Sanz}, \& {Morales}}]{ayalaloera18}
{Ayala-Loera}, C., {Alvarez-Candal}, A., {Ortiz}, J.~L., {et~al.} 2018, \mnras, 481, 1848, \dodoi{10.1093/mnras/sty2363}

\bibitem[{{Bagnulo} {et~al.}(2006){Bagnulo}, {Boehnhardt}, {Muinonen}, {Kolokolova}, {Belskaya}, \& {Barucci}}]{bagnulo06}
{Bagnulo}, S., {Boehnhardt}, H., {Muinonen}, K., {et~al.} 2006, \aap, 450, 1239, \dodoi{10.1051/0004-6361:20054518}

\bibitem[{{Bailey} \& {Malhotra}(2009)}]{bailey09}
{Bailey}, B.~L., \& {Malhotra}, R. 2009, \icarus, 203, 155, \dodoi{10.1016/j.icarus.2009.03.044}

\bibitem[{{Bannister} {et~al.}(2018){Bannister}, {Gladman}, {Kavelaars}, {Petit}, {Volk}, {Chen}, {Alexandersen}, {Gwyn}, {Schwamb}, {Ashton}, {Benecchi}, {Cabral}, {Dawson}, {Delsanti}, {Fraser}, {Granvik}, {Greenstreet}, {Guilbert-Lepoutre}, {Ip}, {Jakubik}, {Jones}, {Kaib}, {Lacerda}, {Van Laerhoven}, {Lawler}, {Lehner}, {Lin}, {Lykawka}, {Marsset}, {Murray-Clay}, {Pike}, {Rousselot}, {Shankman}, {Thirouin}, {Vernazza}, \& {Wang}}]{bannister18}
{Bannister}, M.~T., {Gladman}, B.~J., {Kavelaars}, J.~J., {et~al.} 2018, \apjs, 236, 18, \dodoi{10.3847/1538-4365/aab77a}

\bibitem[{{Barkume} {et~al.}(2008){Barkume}, {Brown}, \& {Schaller}}]{barkume08}
{Barkume}, K.~M., {Brown}, M.~E., \& {Schaller}, E.~L. 2008, \aj, 135, 55, \dodoi{10.1088/0004-6256/135/1/55}

\bibitem[{{Barucci} {et~al.}(2011){Barucci}, {Alvarez-Candal}, {Merlin}, {Belskaya}, {de Bergh}, {Perna}, {DeMeo}, \& {Fornasier}}]{barucci11}
{Barucci}, M.~A., {Alvarez-Candal}, A., {Merlin}, F., {et~al.} 2011, \icarus, 214, 297, \dodoi{10.1016/j.icarus.2011.04.019}

\bibitem[{{Barucci} {et~al.}(2005){Barucci}, {Belskaya}, {Fulchignoni}, \& {Birlan}}]{barucci05}
{Barucci}, M.~A., {Belskaya}, I.~N., {Fulchignoni}, M., \& {Birlan}, M. 2005, \aj, 130, 1291, \dodoi{10.1086/431957}

\bibitem[{{Barucci} {et~al.}(2008){Barucci}, {Brown}, {Emery}, \& {Merlin}}]{barucci08}
{Barucci}, M.~A., {Brown}, M.~E., {Emery}, J.~P., \& {Merlin}, F. 2008, in The Solar System Beyond Neptune, ed. M.~A. {Barucci}, H.~{Boehnhardt}, D.~P. {Cruikshank}, A.~{Morbidelli}, \& R.~{Dotson}, 143--160

\bibitem[{{Barucci} \& {Merlin}(2020)}]{barucci20}
{Barucci}, M.~A., \& {Merlin}, F. 2020, in The Trans-Neptunian Solar System, ed. D.~{Prialnik}, M.~A. {Barucci}, \& L.~{Young}, 109--126, \dodoi{10.1016/B978-0-12-816490-7.00005-9}

\bibitem[{{Bauer} {et~al.}(2013){Bauer}, {Grav}, {Blauvelt}, {Mainzer}, {Masiero}, {Stevenson}, {Kramer}, {Fern{\'a}ndez}, {Lisse}, {Cutri}, {Weissman}, {Dailey}, {Masci}, {Walker}, {Waszczak}, {Nugent}, {Meech}, {Lucas}, {Pearman}, {Wilkins}, {Watkins}, {Kulkarni}, {Wright}, {WISE Team}, \& {PTF Team}}]{bauer13}
{Bauer}, J.~M., {Grav}, T., {Blauvelt}, E., {et~al.} 2013, \apj, 773, 22, \dodoi{10.1088/0004-637X/773/1/22}

\bibitem[{{Belskaya} {et~al.}(2008){Belskaya}, {Levasseur-Regourd}, {Shkuratov}, \& {Muinonen}}]{belskaya08}
{Belskaya}, I.~N., {Levasseur-Regourd}, A.~C., {Shkuratov}, Y.~G., \& {Muinonen}, K. 2008, in The Solar System Beyond Neptune, ed. M.~A. {Barucci}, H.~{Boehnhardt}, D.~P. {Cruikshank}, A.~{Morbidelli}, \& R.~{Dotson}, 115--127

\bibitem[{{Bernardinelli} {et~al.}(2022){Bernardinelli}, {Bernstein}, {Sako}, {Yanny}, {Aguena}, {Allam}, {Andrade-Oliveira}, {Bertin}, {Brooks}, {Buckley-Geer}, {Burke}, {Carnero Rosell}, {Carrasco Kind}, {Carretero}, {Conselice}, {Costanzi}, {da Costa}, {De Vicente}, {Desai}, {Diehl}, {Dietrich}, {Doel}, {Eckert}, {Everett}, {Ferrero}, {Flaugher}, {Fosalba}, {Frieman}, {Garc{\'\i}a-Bellido}, {Gerdes}, {Gruen}, {Gruendl}, {Gschwend}, {Hinton}, {Hollowood}, {Honscheid}, {James}, {Kent}, {Kuehn}, {Kuropatkin}, {Lahav}, {Maia}, {March}, {Menanteau}, {Miquel}, {Morgan}, {Myles}, {Ogando}, {Palmese}, {Paz-Chinch{\'o}n}, {Pieres}, {Plazas Malag{\'o}n}, {Romer}, {Roodman}, {Sanchez}, {Scarpine}, {Schubnell}, {Serrano}, {Sevilla-Noarbe}, {Smith}, {Soares-Santos}, {Suchyta}, {Swanson}, {Tarle}, {To}, {Varga}, \& {Walker}}]{bernardinelli22}
{Bernardinelli}, P.~H., {Bernstein}, G.~M., {Sako}, M., {et~al.} 2022, \apjs, 258, 41, \dodoi{10.3847/1538-4365/ac3914}

\bibitem[{{Bianco} {et~al.}(2022){Bianco}, {Ivezi{\'c}}, {Jones}, {Graham}, {Marshall}, {Saha}, {Strauss}, {Yoachim}, {Ribeiro}, {Anguita}, {Bauer}, {Bauer}, {Bellm}, {Blum}, {Brandt}, {Brough}, {Catelan}, {Clarkson}, {Connolly}, {Gawiser}, {Gizis}, {Hlo{\v{z}}ek}, {Kaviraj}, {Liu}, {Lochner}, {Mahabal}, {Mandelbaum}, {McGehee}, {Neilsen}, {Olsen}, {Peiris}, {Rhodes}, {Richards}, {Ridgway}, {Schwamb}, {Scolnic}, {Shemmer}, {Slater}, {Slosar}, {Smartt}, {Strader}, {Street}, {Trilling}, {Verma}, {Vivas}, {Wechsler}, \& {Willman}}]{bianco22}
{Bianco}, F.~B., {Ivezi{\'c}}, {\v{Z}}., {Jones}, R.~L., {et~al.} 2022, \apjs, 258, 1, \dodoi{10.3847/1538-4365/ac3e72}

\bibitem[{{Brown}(2012)}]{brown12b}
{Brown}, M.~E. 2012, Annual Review of Earth and Planetary Sciences, 40, 467, \dodoi{10.1146/annurev-earth-042711-105352}

\bibitem[{{Buie} {et~al.}(1992){Buie}, {Tholen}, \& {Horne}}]{buie92}
{Buie}, M.~W., {Tholen}, D.~J., \& {Horne}, K. 1992, \icarus, 97, 211, \dodoi{10.1016/0019-1035(92)90129-U}

\bibitem[{{Cabral} {et~al.}(2019){Cabral}, {Guilbert-Lepoutre}, {Fraser}, {Marsset}, {Volk}, {Petit}, {Rousselot}, {Alexandersen}, {Bannister}, {Chen}, {Gladman}, {Gwyn}, \& {Kavelaars}}]{cabral19}
{Cabral}, N., {Guilbert-Lepoutre}, A., {Fraser}, W.~C., {et~al.} 2019, \aap, 621, A102, \dodoi{10.1051/0004-6361/201834021}

\bibitem[{{Chandler} {et~al.}(2024){Chandler}, {Trujillo}, {Oldroyd}, {Kueny}, {Burris}, {Hsieh}, {DeSpain}, {Sedaghat}, {Sheppard}, {Farrell}, {Trilling}, {Gustafsson}, {Magbanua}, {Mazzucato}, {Bosch}, {Shaw-Diaz}, {Gonano}, {Lamperti}, {da Silva Campos}, {Goodwin}, {Terentev}, {Dukes}, \& {Deen}}]{chandler24}
{Chandler}, C.~O., {Trujillo}, C.~A., {Oldroyd}, W.~J., {et~al.} 2024, \aj, 167, 156, \dodoi{10.3847/1538-3881/ad1de2}

\bibitem[{{Chesley} \& {Veres}(2017)}]{chelsey17}
{Chesley}, S.~R., \& {Veres}, P. 2017, arXiv e-prints, arXiv:1705.06209, \dodoi{10.48550/arXiv.1705.06209}

\bibitem[{{Connolly} {et~al.}(2014){Connolly}, {Angeli}, {Chandrasekharan}, {Claver}, {Cook}, {Ivezic}, {Jones}, {Krughoff}, {Peng}, {Peterson}, {Petry}, {Rasmussen}, {Ridgway}, {Saha}, {Sembroski}, {vanderPlas}, \& {Yoachim}}]{connolly14}
{Connolly}, A.~J., {Angeli}, G.~Z., {Chandrasekharan}, S., {et~al.} 2014, in Society of Photo-Optical Instrumentation Engineers (SPIE) Conference Series, Vol. 9150, Modeling, Systems Engineering, and Project Management for Astronomy VI, ed. G.~Z. {Angeli} \& P.~{Dierickx}, 915014, \dodoi{10.1117/12.2054953}

\bibitem[{da~Costa-Luis {et~al.}(2023)da~Costa-Luis, Larroque, Altendorf, Mary, richardsheridan, Korobov, Yorav-Raphael, Ivanov, Bargull, Rodrigues, Chen, Lee, Newey, CrazyPython, JC, Zugnoni, Pagel, mjstevens777, Dektyarev, Rothberg, Plavin, Dill, FichteFoll, Sturm, HeoHeo, van Kemenade, McCracken, MapleCCC, Nordlund, \& Boyle}]{dacostaluis23}
da~Costa-Luis, C., Larroque, S.~K., Altendorf, K., {et~al.} 2023, {tqdm: A fast, Extensible Progress Bar for Python and CLI}, v4.66.1,  Zenodo, \dodoi{10.5281/zenodo.8233425}

\bibitem[{{Delgado} \& {Reuter}(2016)}]{delgago16}
{Delgado}, F., \& {Reuter}, M.~A. 2016, in Society of Photo-Optical Instrumentation Engineers (SPIE) Conference Series, Vol. 9910, Observatory Operations: Strategies, Processes, and Systems VI, ed. A.~B. {Peck}, R.~L. {Seaman}, \& C.~R. {Benn}, 991013, \dodoi{10.1117/12.2233630}

\bibitem[{{Delgado} {et~al.}(2014){Delgado}, {Saha}, {Chandrasekharan}, {Cook}, {Petry}, \& {Ridgway}}]{delgado14}
{Delgado}, F., {Saha}, A., {Chandrasekharan}, S., {et~al.} 2014, in Society of Photo-Optical Instrumentation Engineers (SPIE) Conference Series, Vol. 9150, Modeling, Systems Engineering, and Project Management for Astronomy VI, ed. G.~Z. {Angeli} \& P.~{Dierickx}, 915015, \dodoi{10.1117/12.2056898}

\bibitem[{{Di Sisto} \& {Brunini}(2007)}]{disisto07}
{Di Sisto}, R.~P., \& {Brunini}, A. 2007, \icarus, 190, 224, \dodoi{10.1016/j.icarus.2007.02.012}

\bibitem[{{Di Sisto} \& {Rossignoli}(2020)}]{disisto20}
{Di Sisto}, R.~P., \& {Rossignoli}, N.~L. 2020, Celestial Mechanics and Dynamical Astronomy, 132, 36, \dodoi{10.1007/s10569-020-09971-7}

\bibitem[{{Dobson} {et~al.}(2023){Dobson}, {Schwamb}, {Benecchi}, {Verbiscer}, {Fitzsimmons}, {Shingles}, {Denneau}, {Heinze}, {Smith}, {Tonry}, {Weiland}, \& {Young}}]{dobson23}
{Dobson}, M.~M., {Schwamb}, M.~E., {Benecchi}, S.~D., {et~al.} 2023, \psj, 4, 75, \dodoi{10.3847/PSJ/acc463}

\bibitem[{{Dobson} {et~al.}(2024){Dobson}, {Schwamb}, {Fitzsimmons}, {Schambeau}, {Beck}, {Denneau}, {Erasmus}, {Heinze}, {Shingles}, {Siverd}, {Smith}, {Tonry}, {Weiland}, {Young}, {Kelley}, {Lister}, {Bernardinelli}, {Ferrais}, {Jehin}, {Fedorets}, {Benecchi}, {Verbiscer}, {Murtagh}, {Duffard}, {Gomez}, {Chatelain}, \& {Greenstreet}}]{dobson24}
{Dobson}, M.~M., {Schwamb}, M.~E., {Fitzsimmons}, A., {et~al.} 2024, \psj, 5, 165, \dodoi{10.3847/PSJ/ad543c}

\bibitem[{{Dones} {et~al.}(2015){Dones}, {Brasser}, {Kaib}, \& {Rickman}}]{dones15}
{Dones}, L., {Brasser}, R., {Kaib}, N., \& {Rickman}, H. 2015, \ssr, 197, 191, \dodoi{10.1007/s11214-015-0223-2}

\bibitem[{{Duncan} {et~al.}(2004){Duncan}, {Levison}, \& {Dones}}]{duncan04}
{Duncan}, M., {Levison}, H., \& {Dones}, L. 2004, in Comets II, ed. M.~C. {Festou}, H.~U. {Keller}, \& H.~A. {Weaver}, 193

\bibitem[{{Duncan} \& {Levison}(1997)}]{duncan97}
{Duncan}, M.~J., \& {Levison}, H.~F. 1997, Science, 276, 1670, \dodoi{10.1126/science.276.5319.1670}

\bibitem[{{Elliot} {et~al.}(2005){Elliot}, {Kern}, {Clancy}, {Gulbis}, {Millis}, {Buie}, {Wasserman}, {Chiang}, {Jordan}, {Trilling}, \& {Meech}}]{elliot05}
{Elliot}, J.~L., {Kern}, S.~D., {Clancy}, K.~B., {et~al.} 2005, \aj, 129, 1117, \dodoi{10.1086/427395}

\bibitem[{{Emel'yanenko} {et~al.}(2005){Emel'yanenko}, {Asher}, \& {Bailey}}]{emelyanenko05}
{Emel'yanenko}, V.~V., {Asher}, D.~J., \& {Bailey}, M.~E. 2005, \mnras, 361, 1345, \dodoi{10.1111/j.1365-2966.2005.09267.x}

\bibitem[{{Fedorets} {et~al.}(2020){Fedorets}, {Granvik}, {Jones}, {Juri{\'c}}, \& {Jedicke}}]{fedorets20}
{Fedorets}, G., {Granvik}, M., {Jones}, R.~L., {Juri{\'c}}, M., \& {Jedicke}, R. 2020, \icarus, 338, 113517, \dodoi{10.1016/j.icarus.2019.113517}

\bibitem[{{Fern{\'a}ndez} {et~al.}(2013){Fern{\'a}ndez}, {Kelley}, {Lamy}, {Toth}, {Groussin}, {Lisse}, {A'Hearn}, {Bauer}, {Campins}, {Fitzsimmons}, {Licandro}, {Lowry}, {Meech}, {Pittichov{\'a}}, {Reach}, {Snodgrass}, \& {Weaver}}]{fernandez13}
{Fern{\'a}ndez}, Y.~R., {Kelley}, M.~S., {Lamy}, P.~L., {et~al.} 2013, \icarus, 226, 1138, \dodoi{10.1016/j.icarus.2013.07.021}

\bibitem[{{Fornasier} {et~al.}(2009){Fornasier}, {Barucci}, {de Bergh}, {Alvarez-Candal}, {DeMeo}, {Merlin}, {Perna}, {Guilbert}, {Delsanti}, {Dotto}, \& {Doressoundiram}}]{fornasier09}
{Fornasier}, S., {Barucci}, M.~A., {de Bergh}, C., {et~al.} 2009, \aap, 508, 457, \dodoi{10.1051/0004-6361/200912582}

\bibitem[{{Fornasier} {et~al.}(2014){Fornasier}, {Lazzaro}, {Alvarez-Candal}, {Snodgrass}, {Tozzi}, {Carvano}, {Jim{\'e}nez-Teja}, {Silva}, \& {Bramich}}]{fornasier14}
{Fornasier}, S., {Lazzaro}, D., {Alvarez-Candal}, A., {et~al.} 2014, \aap, 568, L11, \dodoi{10.1051/0004-6361/201424439}

\bibitem[{{Fraser} \& {Brown}(2012)}]{fraser12}
{Fraser}, W.~C., \& {Brown}, M.~E. 2012, \apj, 749, 33, \dodoi{10.1088/0004-637X/749/1/33}

\bibitem[{{Fraser} {et~al.}(2014){Fraser}, {Brown}, {Morbidelli}, {Parker}, \& {Batygin}}]{fraser14}
{Fraser}, W.~C., {Brown}, M.~E., {Morbidelli}, A., {Parker}, A., \& {Batygin}, K. 2014, \apj, 782, 100, \dodoi{10.1088/0004-637X/782/2/100}

\bibitem[{{Fraser} {et~al.}(2022){Fraser}, {Dones}, {Volk}, {Womack}, \& {Nesvorn{\'y}}}]{fraser22}
{Fraser}, W.~C., {Dones}, L., {Volk}, K., {Womack}, M., \& {Nesvorn{\'y}}, D. 2022, arXiv e-prints, arXiv:2210.16354, \dodoi{10.48550/arXiv.2210.16354}

\bibitem[{{Fraser} {et~al.}(2023){Fraser}, {Pike}, {Marsset}, {Schwamb}, {Bannister}, {Buchanan}, {Kavelaars}, {Benecchi}, {Tan}, {Peixinho}, {Gwyn}, {Alexandersen}, {Chen}, {Gladman}, \& {Volk}}]{fraser23}
{Fraser}, W.~C., {Pike}, R.~E., {Marsset}, M., {et~al.} 2023, \psj, 4, 80, \dodoi{10.3847/PSJ/acc844}

\bibitem[{{Gladman} {et~al.}(2008){Gladman}, {Marsden}, \& {Vanlaerhoven}}]{gladman08}
{Gladman}, B., {Marsden}, B.~G., \& {Vanlaerhoven}, C. 2008, in The Solar System Beyond Neptune, ed. M.~A. {Barucci}, H.~{Boehnhardt}, D.~P. {Cruikshank}, A.~{Morbidelli}, \& R.~{Dotson}, 43--57

\bibitem[{{G{'o}rski} {et~al.}(2005){G{'o}rski}, {Hivon}, {Banday}, {Wandelt}, {Hansen}, {Reinecke}, \& {Bartelmann}}]{gorski05}
{G{'o}rski}, K.~M., {Hivon}, E., {Banday}, A.~J., {et~al.} 2005, \apj, 622, 759, \dodoi{10.1086/427976}

\bibitem[{{Grav} {et~al.}(2016){Grav}, {Mainzer}, \& {Spahr}}]{grav16}
{Grav}, T., {Mainzer}, A.~K., \& {Spahr}, T. 2016, \aj, 151, 172, \dodoi{10.3847/0004-6256/151/6/172}

\bibitem[{{Guilbert} {et~al.}(2009){Guilbert}, {Alvarez-Candal}, {Merlin}, {Barucci}, {Dumas}, {de Bergh}, \& {Delsanti}}]{guilbert09}
{Guilbert}, A., {Alvarez-Candal}, A., {Merlin}, F., {et~al.} 2009, \icarus, 201, 272, \dodoi{10.1016/j.icarus.2008.12.023}

\bibitem[{{Guilbert-Lepoutre} {et~al.}(2023){Guilbert-Lepoutre}, {Gkotsinas}, {Raymond}, \& {Nesvorny}}]{guilbertlepoutre23}
{Guilbert-Lepoutre}, A., {Gkotsinas}, A., {Raymond}, S.~N., \& {Nesvorny}, D. 2023, \apj, 942, 92, \dodoi{10.3847/1538-4357/acaa3a}

\bibitem[{{Guy} {et~al.}(2024){Guy}, {Bechtol}, {Bellm}, {Blum}, {Dubois-Felsmann}, {Graham}, {Ivezi\'{c}}, {Lupton}, {Marshall}, {Slater}, \& {Strauss}}]{esp}
{Guy}, L.~P., {Bechtol}, K., {Bellm}, E., {et~al.} 2024, {Rubin Observatory Plans for an Early Science Program}, https://rtn-011.lsst.io/

\bibitem[{Harris {et~al.}(2020)Harris, Millman, van~der Walt, Gommers, Virtanen, Cournapeau, Wieser, Taylor, Berg, Smith, Kern, Picus, Hoyer, van Kerkwijk, Brett, Haldane, del R{\'{i}}o, Wiebe, Peterson, G{\'{e}}rard-Marchant, Sheppard, Reddy, Weckesser, Abbasi, Gohlke, \& Oliphant}]{harris20}
Harris, C.~R., Millman, K.~J., van~der Walt, S.~J., {et~al.} 2020, Nature, 585, 357, \dodoi{10.1038/s41586-020-2649-2}

\bibitem[{{Holman} {et~al.}(2023){Holman}, {Akmal}, {Farnocchia}, {Rein}, {Payne}, {Weryk}, {Tamayo}, \& {Hernandez}}]{holman23}
{Holman}, M.~J., {Akmal}, A., {Farnocchia}, D., {et~al.} 2023, \psj, 4, 69, \dodoi{10.3847/PSJ/acc9a9}

\bibitem[{{Holman} \& {Wisdom}(1993)}]{holman93}
{Holman}, M.~J., \& {Wisdom}, J. 1993, \aj, 105, 1987, \dodoi{10.1086/116574}

\bibitem[{{Holman} {et~al.}(in press){Holman}, {Bernardinelli}, {Schwamb}, {Juri\'{c}}, {Oldag}, {West}, {Merritt}, {Fedorets}, {Cornwall}, {Kurlander}, {Eggl}, {Kubica}, {Kiker}, {Murtagh}, {Jones}, {Yoachim}, {Moeyens}, {Naidu}, \& {Chandler}}]{holman25}
{Holman}, M.~J., {Bernardinelli}, P.~H., {Schwamb}, M.~E., {et~al.} in press

\bibitem[{{Hoover} {et~al.}(2022){Hoover}, {Seligman}, \& {Payne}}]{hoover22}
{Hoover}, D.~J., {Seligman}, D.~Z., \& {Payne}, M.~J. 2022, \psj, 3, 71, \dodoi{10.3847/PSJ/ac58fe}

\bibitem[{{Hsieh} {et~al.}(2021){Hsieh}, {Fitzsimmons}, {Novakovi{\'c}}, {Denneau}, \& {Heinze}}]{hsieh21}
{Hsieh}, H.~H., {Fitzsimmons}, A., {Novakovi{\'c}}, B., {Denneau}, L., \& {Heinze}, A.~N. 2021, \icarus, 354, 114019, \dodoi{10.1016/j.icarus.2020.114019}

\bibitem[{Hunter(2007)}]{hunter07}
Hunter, J.~D. 2007, Computing in Science \& Engineering, 9, 90, \dodoi{10.1109/MCSE.2007.55}

\bibitem[{{Ivezi\'{c}} \& {the LSST Science Collaboration}(2013)}]{ivezic13}
{Ivezi\'{c}}, Z., \& {the LSST Science Collaboration}. 2013, {The LSST System Science Requirements Document}, LSST Science Requirements Document, Cadence Note LPM-17, http://ls.st/LPM-17

\bibitem[{{Ivezi{\'c}} {et~al.}(2019){Ivezi{\'c}}, {Kahn}, {Tyson}, {Abel}, {Acosta}, {Allsman}, {Alonso}, {AlSayyad}, {Anderson}, {Andrew}, {Angel}, {Angeli}, {Ansari}, {Antilogus}, {Araujo}, {Armstrong}, {Arndt}, {Astier}, {Aubourg}, {Auza}, {Axelrod}, {Bard}, {Barr}, {Barrau}, {Bartlett}, {Bauer}, {Bauman}, {Baumont}, {Bechtol}, {Bechtol}, {Becker}, {Becla}, {Beldica}, {Bellavia}, {Bianco}, {Biswas}, {Blanc}, {Blazek}, {Blandford}, {Bloom}, {Bogart}, {Bond}, {Booth}, {Borgland}, {Borne}, {Bosch}, {Boutigny}, {Brackett}, {Bradshaw}, {Brandt}, {Brown}, {Bullock}, {Burchat}, {Burke}, {Cagnoli}, {Calabrese}, {Callahan}, {Callen}, {Carlin}, {Carlson}, {Chandrasekharan}, {Charles-Emerson}, {Chesley}, {Cheu}, {Chiang}, {Chiang}, {Chirino}, {Chow}, {Ciardi}, {Claver}, {Cohen-Tanugi}, {Cockrum}, {Coles}, {Connolly}, {Cook}, {Cooray}, {Covey}, {Cribbs}, {Cui}, {Cutri}, {Daly}, {Daniel}, {Daruich}, {Daubard}, {Daues}, {Dawson}, {Delgado}, {Dellapenna}, {de Peyster}, {de Val-Borro}, {Digel}, {Doherty}, {Dubois},
  {Dubois-Felsmann}, {Durech}, {Economou}, {Eifler}, {Eracleous}, {Emmons}, {Fausti Neto}, {Ferguson}, {Figueroa}, {Fisher-Levine}, {Focke}, {Foss}, {Frank}, {Freemon}, {Gangler}, {Gawiser}, {Geary}, {Gee}, {Geha}, {Gessner}, {Gibson}, {Gilmore}, {Glanzman}, {Glick}, {Goldina}, {Goldstein}, {Goodenow}, {Graham}, {Gressler}, {Gris}, {Guy}, {Guyonnet}, {Haller}, {Harris}, {Hascall}, {Haupt}, {Hernandez}, {Herrmann}, {Hileman}, {Hoblitt}, {Hodgson}, {Hogan}, {Howard}, {Huang}, {Huffer}, {Ingraham}, {Innes}, {Jacoby}, {Jain}, {Jammes}, {Jee}, {Jenness}, {Jernigan}, {Jevremovi{\'c}}, {Johns}, {Johnson}, {Johnson}, {Jones}, {Juramy-Gilles}, {Juri{\'c}}, {Kalirai}, {Kallivayalil}, {Kalmbach}, {Kantor}, {Karst}, {Kasliwal}, {Kelly}, {Kessler}, {Kinnison}, {Kirkby}, {Knox}, {Kotov}, {Krabbendam}, {Krughoff}, {Kub{\'a}nek}, {Kuczewski}, {Kulkarni}, {Ku}, {Kurita}, {Lage}, {Lambert}, {Lange}, {Langton}, {Le Guillou}, {Levine}, {Liang}, {Lim}, {Lintott}, {Long}, {Lopez}, {Lotz}, {Lupton}, {Lust}, {MacArthur}, {Mahabal},
  {Mandelbaum}, {Markiewicz}, {Marsh}, {Marshall}, {Marshall}, {May}, {McKercher}, {McQueen}, {Meyers}, {Migliore}, {Miller}, {Mills}, {Miraval}, {Moeyens}, {Moolekamp}, {Monet}, {Moniez}, {Monkewitz}, {Montgomery}, {Morrison}, {Mueller}, {Muller}, {Mu{\~n}oz Arancibia}, {Neill}, {Newbry}, {Nief}, {Nomerotski}, {Nordby}, {O'Connor}, {Oliver}, {Olivier}, {Olsen}, {O'Mullane}, {Ortiz}, {Osier}, {Owen}, {Pain}, {Palecek}, {Parejko}, {Parsons}, {Pease}, {Peterson}, {Peterson}, {Petravick}, {Libby Petrick}, {Petry}, {Pierfederici}, {Pietrowicz}, {Pike}, {Pinto}, {Plante}, {Plate}, {Plutchak}, {Price}, {Prouza}, {Radeka}, {Rajagopal}, {Rasmussen}, {Regnault}, {Reil}, {Reiss}, {Reuter}, {Ridgway}, {Riot}, {Ritz}, {Robinson}, {Roby}, {Roodman}, {Rosing}, {Roucelle}, {Rumore}, {Russo}, {Saha}, {Sassolas}, {Schalk}, {Schellart}, {Schindler}, {Schmidt}, {Schneider}, {Schneider}, {Schoening}, {Schumacher}, {Schwamb}, {Sebag}, {Selvy}, {Sembroski}, {Seppala}, {Serio}, {Serrano}, {Shaw}, {Shipsey}, {Sick}, {Silvestri},
  {Slater}, {Smith}, {Smith}, {Sobhani}, {Soldahl}, {Storrie-Lombardi}, {Stover}, {Strauss}, {Street}, {Stubbs}, {Sullivan}, {Sweeney}, {Swinbank}, {Szalay}, {Takacs}, {Tether}, {Thaler}, {Thayer}, {Thomas}, {Thornton}, {Thukral}, {Tice}, {Trilling}, {Turri}, {Van Berg}, {Vanden Berk}, {Vetter}, {Virieux}, {Vucina}, {Wahl}, {Walkowicz}, {Walsh}, {Walter}, {Wang}, {Wang}, {Warner}, {Wiecha}, {Willman}, {Winters}, {Wittman}, {Wolff}, {Wood-Vasey}, {Wu}, {Xin}, {Yoachim}, \& {Zhan}}]{ivezic19}
{Ivezi{\'c}}, {\v{Z}}., {Kahn}, S.~M., {Tyson}, J.~A., {et~al.} 2019, \apj, 873, 111, \dodoi{10.3847/1538-4357/ab042c}

\bibitem[{{Jedicke} {et~al.}(2002){Jedicke}, {Larsen}, \& {Spahr}}]{jedicke02}
{Jedicke}, R., {Larsen}, J., \& {Spahr}, T. 2002, in Asteroids III, 71--87

\bibitem[{{Jewitt}(2009)}]{jewitt09}
{Jewitt}, D. 2009, \aj, 137, 4296, \dodoi{10.1088/0004-6256/137/5/4296}

\bibitem[{{Jones} {et~al.}(2020){Jones}, {Yoachim}, {Ivezic}, {Neilsen}, \& {Ribeiro}}]{jones20}
{Jones}, R.~L., {Yoachim}, P., {Ivezic}, Z., {Neilsen}, E.~H., \& {Ribeiro}, T. 2020, {Survey Strategy and Cadence Choices for the Vera C. Rubin Observatory Legacy Survey of Space and Time (LSST)}, v1.2, Zenodo, \dodoi{10.5281/zenodo.4048837}

\bibitem[{{Jones} {et~al.}(2009){Jones}, {Chesley}, {Connolly}, {Harris}, {Ivezic}, {Knezevic}, {Kubica}, {Milani}, \& {Trilling}}]{jones09}
{Jones}, R.~L., {Chesley}, S.~R., {Connolly}, A.~J., {et~al.} 2009, Earth Moon and Planets, 105, 101, \dodoi{10.1007/s11038-009-9305-z}

\bibitem[{{Jones} {et~al.}(2014){Jones}, {Yoachim}, {Chandrasekharan}, {Connolly}, {Cook}, {Ivezic}, {Krughoff}, {Petry}, \& {Ridgway}}]{jones14}
{Jones}, R.~L., {Yoachim}, P., {Chandrasekharan}, S., {et~al.} 2014, in Society of Photo-Optical Instrumentation Engineers (SPIE) Conference Series, Vol. 9149, Observatory Operations: Strategies, Processes, and Systems V, ed. A.~B. {Peck}, C.~R. {Benn}, \& R.~L. {Seaman}, 91490B, \dodoi{10.1117/12.2056835}

\bibitem[{{Jones} {et~al.}(2018){Jones}, {Slater}, {Moeyens}, {Allen}, {Axelrod}, {Cook}, {Ivezi{\'c}}, {Juri{\'c}}, {Myers}, \& {Petry}}]{jones18}
{Jones}, R.~L., {Slater}, C.~T., {Moeyens}, J., {et~al.} 2018, \icarus, 303, 181, \dodoi{10.1016/j.icarus.2017.11.033}

\bibitem[{{Juri\'{c}} {et~al.}(2020){Juri\'{c}}, {Eggl}, {Moeyens}, \& {Jones}}]{juric20}
{Juri\'{c}}, M., {Eggl}, S., {Moeyens}, J., \& {Jones}, R.~L. 2020, {Proposed Modifications to Solar System Processing and Data Products}, Cadence Note DMTN-087, https://dmtn-087.lsst.io/

\bibitem[{Juri\'{c} {et~al.}(2021)Juri\'{c}, Axelrod, Becker, Becla, Bellm, Bosch, Ciardi, Connolly, Dubois-Felsmann, Economou, Freemon, Gelman, Gill, Graham, Guy, Ivezi\'{c}, Jenness, Kantor, Krughoff, Lim, Lupton, Mueller, Nidever, O'Mullane, Patterson, Petravick, Shaw, Slater, Strauss, Swinbank, Tyson, Wood-Vasey, \& Wu}]{juric21}
Juri\'{c}, M., Axelrod, T., Becker, A.~C., {et~al.} 2021, LSST Data Products Definition Document.
\newblock \url{https://lse-163.lsst.io}

\bibitem[{{Kareta} {et~al.}(2020){Kareta}, {Volk}, {Noonan}, {Sharkey}, {Harris}, \& {Reddy}}]{kareta20}
{Kareta}, T., {Volk}, K., {Noonan}, J.~W., {et~al.} 2020, Research Notes of the American Astronomical Society, 4, 74, \dodoi{10.3847/2515-5172/ab963c}

\bibitem[{{Kareta} {et~al.}(2021){Kareta}, {Woodney}, {Schambeau}, {Fernandez}, {Harrington Pinto}, {Wierzchos}, {Womack}, {Bus}, {Steckloff}, {Sarid}, {Volk}, {Harris}, \& {Reddy}}]{kareta21}
{Kareta}, T., {Woodney}, L.~M., {Schambeau}, C., {et~al.} 2021, \psj, 2, 48, \dodoi{10.3847/PSJ/abe23d}

\bibitem[{Kluyver {et~al.}(2016)Kluyver, Ragan-Kelley, P{\'e}rez, Granger, Bussonnier, Frederic, Kelley, Hamrick, Grout, Corlay, Ivanov, Avila, Abdalla, Willing, \& development team}]{kluyver16}
Kluyver, T., Ragan-Kelley, B., P{\'e}rez, F., {et~al.} 2016, in Positioning and Power in Academic Publishing: Players, Agents and Agendas (IOS Press), 87--90.
\newblock \url{https://eprints.soton.ac.uk/403913/}

\bibitem[{{Kulyk} {et~al.}(2016){Kulyk}, {Korsun}, {Rousselot}, {Afanasiev}, \& {Ivanova}}]{kulyk16}
{Kulyk}, I., {Korsun}, P., {Rousselot}, P., {Afanasiev}, V., \& {Ivanova}, O. 2016, \icarus, 271, 314, \dodoi{10.1016/j.icarus.2016.01.037}

\bibitem[{{Kurlander} {et~al.}(2025){Kurlander}, {Holman}, {Bernardinelli}, {Juri{\'c}}, {Heinze}, \& {Payne}}]{kurlander24}
{Kurlander}, J.~A., {Holman}, M.~J., {Bernardinelli}, P.~H., {et~al.} 2025, \aj, 169, 73, \dodoi{10.3847/1538-3881/ad9a58}

\bibitem[{{Kurucz}(2005)}]{kurucz05}
{Kurucz}, R.~L. 2005, Memorie della Societa Astronomica Italiana Supplementi, 8, 189

\bibitem[{{Lam} {et~al.}(2015){Lam}, {Pitrou}, \& {Seibert}}]{lam15}
{Lam}, S.~K., {Pitrou}, A., \& {Seibert}, S. 2015, in Proc. Second Workshop on the LLVM Compiler Infrastructure in HPC, 1--6, \dodoi{10.1145/2833157.2833162}

\bibitem[{{Lawler} {et~al.}(2018{\natexlab{a}}){Lawler}, {Kavelaars}, {Alexandersen}, {Bannister}, {Gladman}, {Petit}, \& {Shankman}}]{lawler18b}
{Lawler}, S.~M., {Kavelaars}, J.~J., {Alexandersen}, M., {et~al.} 2018{\natexlab{a}}, Frontiers in Astronomy and Space Sciences, 5, 14, \dodoi{10.3389/fspas.2018.00014}

\bibitem[{{Lawler} {et~al.}(2018{\natexlab{b}}){Lawler}, {Shankman}, {Kavelaars}, {Alexandersen}, {Bannister}, {Chen}, {Gladman}, {Fraser}, {Gwyn}, {Kaib}, {Petit}, \& {Volk}}]{lawler18}
{Lawler}, S.~M., {Shankman}, C., {Kavelaars}, J.~J., {et~al.} 2018{\natexlab{b}}, \aj, 155, 197, \dodoi{10.3847/1538-3881/aab8ff}

\bibitem[{{Levison} \& {Duncan}(1997)}]{levison97}
{Levison}, H.~F., \& {Duncan}, M.~J. 1997, \icarus, 127, 13, \dodoi{10.1006/icar.1996.5637}

\bibitem[{{Lilly} {et~al.}(2024){Lilly}, {Jev{\v{c}}{\'a}k}, {Schambeau}, {Volk}, {Steckloff}, {Hsieh}, {Fernandez}, {Bauer}, {Weryk}, \& {Wainscoat}}]{lilly24}
{Lilly}, E., {Jev{\v{c}}{\'a}k}, P., {Schambeau}, C., {et~al.} 2024, \apjl, 960, L8, \dodoi{10.3847/2041-8213/ad1606}

\bibitem[{{LSST Science Collaboration} {et~al.}(2009){LSST Science Collaboration}, Abell, Allison, Anderson, Andrew, Angel, Armus, Arnett, Asztalos, Axelrod, Bailey, Ballantyne, Bankert, Barkhouse, Barr, Barrientos, Barth, Bartlett, Becker, Becla, Beers, Bernstein, Biswas, Blanton, Bloom, Bochanski, Boeshaar, Borne, Bradac, Brandt, Bridge, Brown, Brunner, Bullock, Burgasser, Burge, Burke, Cargile, Chandrasekharan, Chartas, Chesley, Chu, Cinabro, Claire, Claver, Clowe, Connolly, Cook, Cooke, Cooray, Covey, Culliton, de~Jong, de~Vries, Debattista, Delgado, Dell'Antonio, Dhital, Stefano, Dickinson, Dilday, Djorgovski, Dobler, Donalek, Dubois-Felsmann, Durech, Eliasdottir, Eracleous, Eyer, Falco, Fan, Fassnacht, Ferguson, Fernandez, Fields, Finkbeiner, Figueroa, Fox, Francke, Frank, Frieman, Fromenteau, Furqan, Galaz, Gal-Yam, Garnavich, Gawiser, Geary, Gee, Gibson, Gilmore, Grace, Green, Gressler, Grillmair, Habib, Haggerty, Hamuy, Harris, Hawley, Heavens, Hebb, Henry, Hileman, Hilton, Hoadley, Holberg, Holman,
  Howell, Infante, Ivezic, Jacoby, Jain, R, Jedicke, Jee, Jernigan, Jha, Johnston, Jones, Juric, Kaasalainen, Styliani, Kafka, Kahn, Kaib, Kalirai, Kantor, Kasliwal, Keeton, Kessler, Knezevic, Kowalski, Krabbendam, Krughoff, Kulkarni, Kuhlman, Lacy, Lepine, Liang, Lien, Lira, Long, Lorenz, Lotz, Lupton, Lutz, Macri, Mahabal, Mandelbaum, Marshall, May, McGehee, Meadows, Meert, Milani, Miller, Miller, Mills, Minniti, Monet, Mukadam, Nakar, Neill, Newman, Nikolaev, Nordby, O'Connor, Oguri, Oliver, Olivier, Olsen, Olsen, Olszewski, Oluseyi, Padilla, Parker, Pepper, Peterson, Petry, Pinto, Pizagno, Popescu, Prsa, Radcka, Raddick, Rasmussen, Rau, Rho, Rhoads, Richards, Ridgway, Robertson, Roskar, Saha, Sarajedini, Scannapieco, Schalk, Schindler, Schmidt, Schmidt, Schneider, Schumacher, Scranton, Sebag, Seppala, Shemmer, Simon, Sivertz, Smith, Smith, Smith, Spitz, Stanford, Stassun, Strader, Strauss, Stubbs, Sweeney, Szalay, Szkody, Takada, Thorman, Trilling, Trimble, Tyson, Berg, Berk, VanderPlas, Verde, Vrsnak,
  Walkowicz, Wandelt, Wang, Wang, Warner, Wechsler, West, Wiecha, Williams, Willman, Wittman, Wolff, Wood-Vasey, Wozniak, Young, Zentner, \& Zhan}]{lsst09}
{LSST Science Collaboration}, Abell, P.~A., Allison, J., {et~al.} 2009, LSST Science Book, Version 2.0.
\newblock \doarXiv{0912.0201}

\bibitem[{{LSST Science Collaboration} {et~al.}(2017){LSST Science Collaboration}, {Marshall}, {Anguita}, {Bianco}, {Bellm}, {Brandt}, {Clarkson}, {Connolly}, {Gawiser}, {Ivezic}, {Jones}, {Lochner}, {Lund}, {Mahabal}, {Nidever}, {Olsen}, {Ridgway}, {Rhodes}, {Shemmer}, {Trilling}, {Vivas}, {Walkowicz}, {Willman}, {Yoachim}, {Anderson}, {Antilogus}, {Angus}, {Arcavi}, {Awan}, {Biswas}, {Bell}, {Bennett}, {Britt}, {Buzasi}, {Casetti-Dinescu}, {Chomiuk}, {Claver}, {Cook}, {Davenport}, {Debattista}, {Digel}, {Doctor}, {Firth}, {Foley}, {Fong}, {Galbany}, {Giampapa}, {Gizis}, {Graham}, {Grillmair}, {Gris}, {Haiman}, {Hartigan}, {Hawley}, {Hlozek}, {Jha}, {Johns-Krull}, {Kanbur}, {Kalogera}, {Kashyap}, {Kasliwal}, {Kessler}, {Kim}, {Kurczynski}, {Lahav}, {Liu}, {Malz}, {Margutti}, {Matheson}, {McEwen}, {McGehee}, {Meibom}, {Meyers}, {Monet}, {Neilsen}, {Newman}, {O'Dowd}, {Peiris}, {Penny}, {Peters}, {Poleski}, {Ponder}, {Richards}, {Rho}, {Rubin}, {Schmidt}, {Schuhmann}, {Shporer}, {Slater}, {Smith},
  {Soares-Santos}, {Stassun}, {Strader}, {Strauss}, {Street}, {Stubbs}, {Sullivan}, {Szkody}, {Trimble}, {Tyson}, {de Val-Borro}, {Valenti}, {Wagoner}, {Wood-Vasey}, \& {Zauderer}}]{lsst17}
{LSST Science Collaboration}, {Marshall}, P., {Anguita}, T., {et~al.} 2017, arXiv e-prints, arXiv:1708.04058, \dodoi{10.48550/arXiv.1708.04058}

\bibitem[{{Mazzotta Epifani} {et~al.}(2018){Mazzotta Epifani}, {Dotto}, {Ieva}, {Perna}, {Palumbo}, {Micheli}, \& {Perozzi}}]{mazzotta18}
{Mazzotta Epifani}, E., {Dotto}, E., {Ieva}, S., {et~al.} 2018, \aap, 620, A93, \dodoi{10.1051/0004-6361/201731224}

\bibitem[{{Merritt} {et~al.}(in press){Merritt}, {Fedorets}, {Schwamb}, {Cornwall}, {Bernardinelli}, \& {Juri\'{c}}}]{merritt25}
{Merritt}, S.~R., {Fedorets}, G., {Schwamb}, M.~E., {et~al.} in press

\bibitem[{{Mommert} {et~al.}(2019){Mommert}, {Kelley}, {de Val-Borro}, {Li}, {Guzman}, {Sip{\H{o}}cz}, {{\v{D}}urech}, {Granvik}, {Grundy}, {Moskovitz}, {Penttil{\"a}}, \& {Samarasinha}}]{mommert19}
{Mommert}, M., {Kelley}, M., {de Val-Borro}, M., {et~al.} 2019, The Journal of Open Source Software, 4, 1426, \dodoi{10.21105/joss.01426}

\bibitem[{{Myers} {et~al.}(2013){Myers}, {Jones}, \& {Axelrod}}]{myers13}
{Myers}, J., {Jones}, R.~L., \& {Axelrod}, T. 2013, {Moving Object Pipeline System Design}, Cadence Note LDM-156, https://docushare.lsst.org/docushare/ dsweb/Get/LDM-156/LDM-156.pdf

\bibitem[{{Naghib} {et~al.}(2019){Naghib}, {Yoachim}, {Vanderbei}, {Connolly}, \& {Jones}}]{naghib19}
{Naghib}, E., {Yoachim}, P., {Vanderbei}, R.~J., {Connolly}, A.~J., \& {Jones}, R.~L. 2019, \aj, 157, 151, \dodoi{10.3847/1538-3881/aafece}

\bibitem[{{Nesvorn{\'y}} {et~al.}(2019){Nesvorn{\'y}}, {Vokrouhlick{\'y}}, {Stern}, {Davidsson}, {Bannister}, {Volk}, {Chen}, {Gladman}, {Kavelaars}, {Petit}, {Gwyn}, \& {Alexandersen}}]{nesvorny19}
{Nesvorn{\'y}}, D., {Vokrouhlick{\'y}}, D., {Stern}, A.~S., {et~al.} 2019, \aj, 158, 132, \dodoi{10.3847/1538-3881/ab3651}

\bibitem[{pandas~development team(2020)}]{pandas20}
pandas~development team, T. 2020, pandas-dev/pandas: Pandas, latest,  Zenodo, \dodoi{10.5281/zenodo.3509134}

\bibitem[{{Peixinho} {et~al.}(2012){Peixinho}, {Delsanti}, {Guilbert-Lepoutre}, {Gafeira}, \& {Lacerda}}]{peixinho12}
{Peixinho}, N., {Delsanti}, A., {Guilbert-Lepoutre}, A., {Gafeira}, R., \& {Lacerda}, P. 2012, \aap, 546, A86, \dodoi{10.1051/0004-6361/201219057}

\bibitem[{{Peixinho} {et~al.}(2003){Peixinho}, {Doressoundiram}, {Delsanti}, {Boehnhardt}, {Barucci}, \& {Belskaya}}]{peixinho03}
{Peixinho}, N., {Doressoundiram}, A., {Delsanti}, A., {et~al.} 2003, \aap, 410, L29, \dodoi{10.1051/0004-6361:20031420}

\bibitem[{{Peixinho} {et~al.}(2020){Peixinho}, {Thirouin}, {Tegler}, {Di Sisto}, {Delsanti}, {Guilbert-Lepoutre}, \& {Bauer}}]{peixinho20}
{Peixinho}, N., {Thirouin}, A., {Tegler}, S.~C., {et~al.} 2020, in The Trans-Neptunian Solar System, ed. D.~{Prialnik}, M.~A. {Barucci}, \& L.~{Young}, 307--329, \dodoi{10.1016/B978-0-12-816490-7.00014-X}

\bibitem[{{Perna} {et~al.}(2010){Perna}, {Barucci}, {Fornasier}, {DeMeo}, {Alvarez-Candal}, {Merlin}, {Dotto}, {Doressoundiram}, \& {de Bergh}}]{perna10}
{Perna}, D., {Barucci}, M.~A., {Fornasier}, S., {et~al.} 2010, \aap, 510, A53, \dodoi{10.1051/0004-6361/200913654}

\bibitem[{{Petit} {et~al.}(2008){Petit}, {Kavelaars}, {Gladman}, \& {Loredo}}]{petit08}
{Petit}, J.~M., {Kavelaars}, J.~J., {Gladman}, B., \& {Loredo}, T. 2008, in The Solar System Beyond Neptune, ed. M.~A. {Barucci}, H.~{Boehnhardt}, D.~P. {Cruikshank}, A.~{Morbidelli}, \& R.~{Dotson}, 71--87

\bibitem[{{Prialnik}(1992)}]{prialnik92}
{Prialnik}, D. 1992, \apj, 388, 196, \dodoi{10.1086/171143}

\bibitem[{{PyTables Developers Team}(2002)}]{pytables02}
{PyTables Developers Team}. 2002, {PyTables}: Hierarchical Datasets in {Python}.
\newblock \url{https://www.pytables.org/}

\bibitem[{{Rabinowitz} {et~al.}(2012){Rabinowitz}, {Schwamb}, {Hadjiyska}, \& {Tourtellotte}}]{rabinowitz12}
{Rabinowitz}, D., {Schwamb}, M.~E., {Hadjiyska}, E., \& {Tourtellotte}, S. 2012, \aj, 144, 140, \dodoi{10.1088/0004-6256/144/5/140}

\bibitem[{{Rabinowitz} {et~al.}(2007){Rabinowitz}, {Schaefer}, \& {Tourtellotte}}]{rabinowitz07}
{Rabinowitz}, D.~L., {Schaefer}, B.~E., \& {Tourtellotte}, S.~W. 2007, \aj, 133, 26, \dodoi{10.1086/508931}

\bibitem[{Rein {et~al.}(2023)Rein, Holman, \& Akmal}]{rein23}
Rein, H., Holman, M., \& Akmal, A. 2023, matthewholman/assist: v1.1.1, v1.1.1,  Zenodo, \dodoi{10.5281/zenodo.7778017}

\bibitem[{{Rein} \& {Liu}(2012)}]{rein12}
{Rein}, H., \& {Liu}, S.~F. 2012, \aap, 537, A128, \dodoi{10.1051/0004-6361/201118085}

\bibitem[{{Rein} \& {Spiegel}(2015)}]{rein15}
{Rein}, H., \& {Spiegel}, D.~S. 2015, \mnras, 446, 1424, \dodoi{10.1093/mnras/stu2164}

\bibitem[{{Robinson} {et~al.}(submitted){Robinson}, {Schwamb}, {Jones}, {Juri\'{c}}, {Yoachim}, {Parajeko}, R., \& {Fraser}}]{robinson24b}
{Robinson}, J.~E., {Schwamb}, M.~E., {Jones}, R.~L., {et~al.} submitted

\bibitem[{{Robinson} {et~al.}(2024){Robinson}, {Fitzsimmons}, {Young}, {Bannister}, {Denneau}, {Erasmus}, {Lawrence}, {Siverd}, \& {Tonry}}]{robinson24}
{Robinson}, J.~E., {Fitzsimmons}, A., {Young}, D.~R., {et~al.} 2024, \mnras, 531, 304, \dodoi{10.1093/mnras/stae966}

\bibitem[{{Rousselot} {et~al.}(2005){Rousselot}, {Petit}, {Poulet}, \& {Sergeev}}]{rousselot05}
{Rousselot}, P., {Petit}, J.~M., {Poulet}, F., \& {Sergeev}, A. 2005, \icarus, 176, 478, \dodoi{10.1016/j.icarus.2005.03.001}

\bibitem[{{Sarid} {et~al.}(2019){Sarid}, {Volk}, {Steckloff}, {Harris}, {Womack}, \& {Woodney}}]{sarid19}
{Sarid}, G., {Volk}, K., {Steckloff}, J.~K., {et~al.} 2019, \apjl, 883, L25, \dodoi{10.3847/2041-8213/ab3fb3}

\bibitem[{{Schaefer} {et~al.}(2009){Schaefer}, {Rabinowitz}, \& {Tourtellotte}}]{schaefer09}
{Schaefer}, B.~E., {Rabinowitz}, D.~L., \& {Tourtellotte}, S.~W. 2009, \aj, 137, 129, \dodoi{10.1088/0004-6256/137/1/129}

\bibitem[{{Schwamb} {et~al.}(2018){Schwamb}, {Jones}, {Chesley}, {Fitzsimmons}, {Fraser}, {Holman}, {Hsieh}, {Ragozzine}, {Thomas}, {Trilling}, {Brown}, {Bannister}, {Bodewits}, {de Val-Borro}, {Gerdes}, {Granvik}, {Kelley}, {Knight}, {Seaman}, {Ye}, \& {Young}}]{schwamb18}
{Schwamb}, M.~E., {Jones}, R.~L., {Chesley}, S.~R., {et~al.} 2018, arXiv e-prints, arXiv:1802.01783, \dodoi{10.48550/arXiv.1802.01783}

\bibitem[{{Schwamb} {et~al.}(2019){Schwamb}, {Fraser}, {Bannister}, {Marsset}, {Pike}, {Kavelaars}, {Benecchi}, {Lehner}, {Wang}, {Thirouin}, {Delsanti}, {Peixinho}, {Volk}, {Alexandersen}, {Chen}, {Gladman}, {Gwyn}, \& {Petit}}]{schwamb19}
{Schwamb}, M.~E., {Fraser}, W.~C., {Bannister}, M.~T., {et~al.} 2019, \apjs, 243, 12, \dodoi{10.3847/1538-4365/ab2194}

\bibitem[{{Schwamb} {et~al.}(2021){Schwamb}, {Juri{\'c}}, {Bolin}, {Dones}, {Greenstreet}, {Hsieh}, {Inno}, {Jones}, {Kelley}, {Knight}, {Reach}, {Seccull}, {Snodgrass}, {Trilling}, \& {Vera C. Rubin Observatory LSST Solar System Science Collaboration}}]{schwamb21}
{Schwamb}, M.~E., {Juri{\'c}}, M., {Bolin}, B.~T., {et~al.} 2021, Research Notes of the American Astronomical Society, 5, 143, \dodoi{10.3847/2515-5172/ac090f}

\bibitem[{{Schwamb} {et~al.}(2023){Schwamb}, {Jones}, {Yoachim}, {Volk}, {Dorsey}, {Opitom}, {Greenstreet}, {Lister}, {Snodgrass}, {Bolin}, {Inno}, {Bannister}, {Eggl}, {Solontoi}, {Kelley}, {Juri{\'c}}, {Lin}, {Ragozzine}, {Bernardinelli}, {Chesley}, {Daylan}, {{\v{D}}urech}, {Fraser}, {Granvik}, {Knight}, {Lisse}, {Malhotra}, {Oldroyd}, {Thirouin}, \& {Ye}}]{schwamb23}
{Schwamb}, M.~E., {Jones}, R.~L., {Yoachim}, P., {et~al.} 2023, \apjs, 266, 22, \dodoi{10.3847/1538-4365/acc173}

\bibitem[{{SCOC}(2022)}]{scocv1}
{SCOC}. 2022, {Survey Cadence Optimization Committee's Phase 1 Recommendations}, Cadence Note PSTN-056, https://pstn-053.lsst.io/

\bibitem[{{SCOC}(2023)}]{scocv2}
---. 2023, {Survey Cadence Optimization Committee's Phase 2 Recommendations}, Cadence Note PSTN-056, https://pstn-055.lsst.io/

\bibitem[{{SCOC}(2024)}]{scocv3}
---. 2024, {Survey Cadence Optimization Committee's Phase 3 Recommendations}, Cadence Note PSTN-056, https://pstn-056.lsst.io/

\bibitem[{{Seccull} {et~al.}(2019){Seccull}, {Fraser}, {Puzia}, {Fitzsimmons}, \& {Cupani}}]{seccull19}
{Seccull}, T., {Fraser}, W.~C., {Puzia}, T.~H., {Fitzsimmons}, A., \& {Cupani}, G. 2019, \aj, 157, 88, \dodoi{10.3847/1538-3881/aafbe4}

\bibitem[{{Seligman} {et~al.}(2021){Seligman}, {Kratter}, {Levine}, \& {Jedicke}}]{seligman21}
{Seligman}, D.~Z., {Kratter}, K.~M., {Levine}, W.~G., \& {Jedicke}, R. 2021, \psj, 2, 234, \dodoi{10.3847/PSJ/ac2dee}

\bibitem[{{Shankman} {et~al.}(2013){Shankman}, {Gladman}, {Kaib}, {Kavelaars}, \& {Petit}}]{shankman13}
{Shankman}, C., {Gladman}, B.~J., {Kaib}, N., {Kavelaars}, J.~J., \& {Petit}, J.~M. 2013, \apjl, 764, L2, \dodoi{10.1088/2041-8205/764/1/L2}

\bibitem[{{Shankman} {et~al.}(2016){Shankman}, {Kavelaars}, {Gladman}, {Alexandersen}, {Kaib}, {Petit}, {Bannister}, {Chen}, {Gwyn}, {Jakubik}, \& {Volk}}]{shankman16}
{Shankman}, C., {Kavelaars}, J., {Gladman}, B.~J., {et~al.} 2016, \aj, 151, 31, \dodoi{10.3847/0004-6256/151/2/31}

\bibitem[{{Shannon} {et~al.}(2015){Shannon}, {Jackson}, {Veras}, \& {Wyatt}}]{shannon15}
{Shannon}, A., {Jackson}, A.~P., {Veras}, D., \& {Wyatt}, M. 2015, \mnras, 446, 2059, \dodoi{10.1093/mnras/stu2267}

\bibitem[{{Sheppard} {et~al.}(2000){Sheppard}, {Jewitt}, {Trujillo}, {Brown}, \& {Ashley}}]{sheppard00}
{Sheppard}, S.~S., {Jewitt}, D.~C., {Trujillo}, C.~A., {Brown}, M. J.~I., \& {Ashley}, M. C.~B. 2000, \aj, 120, 2687, \dodoi{10.1086/316805}

\bibitem[{{Sheppard} {et~al.}(2016){Sheppard}, {Trujillo}, \& {Tholen}}]{sheppard16}
{Sheppard}, S.~S., {Trujillo}, C., \& {Tholen}, D.~J. 2016, \apjl, 825, L13, \dodoi{10.3847/2041-8205/825/1/L13}

\bibitem[{{Sheppard} {et~al.}(2011){Sheppard}, {Udalski}, {Trujillo}, {Kubiak}, {Pietrzynski}, {Poleski}, {Soszynski}, {Szyma{\'n}ski}, \& {Ulaczyk}}]{sheppard11}
{Sheppard}, S.~S., {Udalski}, A., {Trujillo}, C., {et~al.} 2011, \aj, 142, 98, \dodoi{10.1088/0004-6256/142/4/98}

\bibitem[{{Silsbee} \& {Tremaine}(2016)}]{silsbee16}
{Silsbee}, K., \& {Tremaine}, S. 2016, \aj, 152, 103, \dodoi{10.3847/0004-6256/152/4/103}

\bibitem[{{Solontoi} {et~al.}(2010){Solontoi}, {Ivezi{\'c}}, {West}, {Claire}, {Juri{\'c}}, {Becker}, {Jones}, {Hall}, {Kent}, {Lupton}, {Knapp}, {Quinn}, {Gunn}, {Schneider}, \& {Loomis}}]{solontoi10}
{Solontoi}, M., {Ivezi{\'c}}, {\v{Z}}., {West}, A.~A., {et~al.} 2010, \icarus, 205, 605, \dodoi{10.1016/j.icarus.2009.07.042}

\bibitem[{{Steckloff} {et~al.}(2020){Steckloff}, {Sarid}, {Volk}, {Kareta}, {Womack}, {Harris}, {Woodney}, \& {Schambeau}}]{steckloff20}
{Steckloff}, J.~K., {Sarid}, G., {Volk}, K., {et~al.} 2020, \apjl, 904, L20, \dodoi{10.3847/2041-8213/abc888}

\bibitem[{{Tegler} {et~al.}(2008){Tegler}, {Bauer}, {Romanishin}, \& {Peixinho}}]{tegler08}
{Tegler}, S.~C., {Bauer}, J.~M., {Romanishin}, W., \& {Peixinho}, N. 2008, in The Solar System Beyond Neptune, ed. M.~A. {Barucci}, H.~{Boehnhardt}, D.~P. {Cruikshank}, A.~{Morbidelli}, \& R.~{Dotson}, 105--114

\bibitem[{{Tegler} \& {Romanishin}(1998)}]{tegler98}
{Tegler}, S.~C., \& {Romanishin}, W. 1998, \nat, 392, 49, \dodoi{10.1038/32108}

\bibitem[{{Tegler} \& {Romanishin}(2000)}]{tegler00}
---. 2000, \nat, 407, 979, \dodoi{10.1038/35039572}

\bibitem[{{Tegler} \& {Romanishin}(2003)}]{tegler03}
---. 2003, \icarus, 161, 181, \dodoi{10.1016/S0019-1035(02)00021-0}

\bibitem[{{Tegler} {et~al.}(2016){Tegler}, {Romanishin}, {Consolmagno}, \& {J.}}]{tegler16}
{Tegler}, S.~C., {Romanishin}, W., {Consolmagno}, G.~J., \& {J.}, S. 2016, \aj, 152, 210, \dodoi{10.3847/0004-6256/152/6/210}

\bibitem[{{Tiscareno} \& {Malhotra}(2003)}]{tiscareno03}
{Tiscareno}, M.~S., \& {Malhotra}, R. 2003, \aj, 126, 3122, \dodoi{10.1086/379554}

\bibitem[{{Trujillo}(2008)}]{trujillo08}
{Trujillo}, C.~A. 2008, in The Solar System Beyond Neptune, ed. M.~A. {Barucci}, H.~{Boehnhardt}, D.~P. {Cruikshank}, A.~{Morbidelli}, \& R.~{Dotson}, 573--585

\bibitem[{Uieda {et~al.}(2020)Uieda, Soler, Rampin, van Kemenade, Turk, Shapero, Banihirwe, \& Leeman}]{uieda20}
Uieda, L., Soler, S., Rampin, R., {et~al.} 2020, Journal of Open Source Software, 5, 1943, \dodoi{10.21105/joss.01943}

\bibitem[{{van der Velden}(2020)}]{vandervelden20}
{van der Velden}, E. 2020, The Journal of Open Source Software, 5, 2004, \dodoi{10.21105/joss.02004}

\bibitem[{{Verbiscer} {et~al.}(2013){Verbiscer}, {Helfenstein}, \& {Buratti}}]{verbiscer13}
{Verbiscer}, A.~J., {Helfenstein}, P., \& {Buratti}, B.~J. 2013, in Astrophysics and Space Science Library, Vol. 356, Astrophysics and Space Science Library, ed. M.~S. {Gudipati} \& J.~{Castillo-Rogez}, 47, \dodoi{10.1007/978-1-4614-3076-6_2}

\bibitem[{{Vere{\v{s}}} \& {Chesley}(2017)}]{veres17}
{Vere{\v{s}}}, P., \& {Chesley}, S.~R. 2017, \aj, 154, 13, \dodoi{10.3847/1538-3881/aa73d0}

\bibitem[{{Virtanen} {et~al.}(2020){Virtanen}, {Gommers}, {Oliphant}, {Haberland}, {Reddy}, {Cournapeau}, {Burovski}, {Peterson}, {Weckesser}, {Bright}, {van der Walt}, {Brett}, {Wilson}, {Millman}, {Mayorov}, {Nelson}, {Jones}, {Kern}, {Larson}, {Carey}, {Polat}, {Feng}, {Moore}, {VanderPlas}, {Laxalde}, {Perktold}, {Cimrman}, {Henriksen}, {Quintero}, {Harris}, {Archibald}, {Ribeiro}, {Pedregosa}, {van Mulbregt}, \& {SciPy 1. 0 Contributors}}]{virtanen20}
{Virtanen}, P., {Gommers}, R., {Oliphant}, T.~E., {et~al.} 2020, Nature Methods, 17, 261, \dodoi{10.1038/s41592-019-0686-2}

\bibitem[{{Volk} \& {Malhotra}(2008)}]{volk08}
{Volk}, K., \& {Malhotra}, R. 2008, \apj, 687, 714, \dodoi{10.1086/591839}

\bibitem[{{Volk} \& {Van Laerhoven}(2024)}]{volk24}
{Volk}, K., \& {Van Laerhoven}, C. 2024, Research Notes of the American Astronomical Society, 8, 36, \dodoi{10.3847/2515-5172/ad22d4}

\bibitem[{Waskom(2021)}]{waskom21}
Waskom, M.~L. 2021, Journal of Open Source Software, 6, 3021, \dodoi{10.21105/joss.03021}

\bibitem[{{Weryk} {et~al.}(2016){Weryk}, {Lilly}, {Chastel}, {Denneau}, {Jedicke}, {Magnier}, {Wainscoat}, {Chambers}, {Flewelling}, {Huber}, {Waters}, \& {PS1 Builders}}]{weryk16}
{Weryk}, R.~J., {Lilly}, E., {Chastel}, S., {et~al.} 2016, arXiv e-prints, arXiv:1607.04895, \dodoi{10.48550/arXiv.1607.04895}

\bibitem[{{W}es {M}c{K}inney(2010)}]{mckinney10}
{W}es {M}c{K}inney. 2010, in {P}roceedings of the 9th {P}ython in {S}cience {C}onference, ed. {S}t\'efan van~der {W}alt \& {J}arrod {M}illman, 56 -- 61, \dodoi{10.25080/Majora-92bf1922-00a}

\bibitem[{{Wong} \& {Brown}(2016)}]{wong16}
{Wong}, I., \& {Brown}, M.~E. 2016, \aj, 152, 90, \dodoi{10.3847/0004-6256/152/4/90}

\bibitem[{{Wong} \& {Brown}(2017)}]{wong17}
---. 2017, \aj, 153, 145, \dodoi{10.3847/1538-3881/aa60c3}

\bibitem[{Yoachim {et~al.}(2024)Yoachim, Jones, Eric H.~Neilsen, Bechtol, Becker, \& Ross}]{yoachim24b}
Yoachim, P., Jones, L., Eric H.~Neilsen, J., {et~al.} 2024, lsst/rubin\_scheduler: v3.4.0, v3.4.0,  Zenodo, \dodoi{10.5281/zenodo.14232232}

\bibitem[{{Yoachim} {et~al.}(2016){Yoachim}, {Coughlin}, {Angeli}, {Claver}, {Connolly}, {Cook}, {Daniel}, {Ivezi{\'c}}, {Jones}, {Petry}, {Reuter}, {Stubbs}, \& {Xin}}]{yoachim16}
{Yoachim}, P., {Coughlin}, M., {Angeli}, G.~Z., {et~al.} 2016, in Society of Photo-Optical Instrumentation Engineers (SPIE) Conference Series, Vol. 9910, Observatory Operations: Strategies, Processes, and Systems VI, ed. A.~B. {Peck}, R.~L. {Seaman}, \& C.~R. {Benn}, 99101A, \dodoi{10.1117/12.2232947}

\bibitem[{{Yoachim} {et~al.}(2023){Yoachim}, {Jones}, {Eric H. Neilsen}, {Tiago}, {Parejko}, {Carlin}, {Becker}, {pgris}, {Prisinzano}, {Dennihy}, {Bellm}, {Sick}, {lmptc}, {LI}, {nsabrams}, {Guy}, {Bricman}, {Bregeon}, {Lim}, {Kelley}, \& {Andreoni}}]{yoachim23}
{Yoachim}, P., {Jones}, L., {Eric H. Neilsen}, J., {et~al.} 2023, {lsst/rubin\_sim: v2.0.0}, v2.0.0,  Zenodo, \dodoi{10.5281/zenodo.10215451}

\bibitem[{Zonca {et~al.}(2019)Zonca, Singer, Lenz, Reinecke, Rosset, Hivon, \& Gorski}]{zonca19}
Zonca, A., Singer, L., Lenz, D., {et~al.} 2019, Journal of Open Source Software, 4, 1298, \dodoi{10.21105/joss.01298}

\end{thebibliography}
\bibliographystyle{aasjournal}



\end{document}